\documentclass[12pt, draftclsnofoot, onecolumn]{IEEEtran}

\usepackage{amssymb,amsmath,float,bm}
\usepackage{subfig}
\usepackage[pdftex]{graphicx}
\usepackage[skip=1pt]{caption}
\graphicspath{{./Figures/}}
\usepackage{epstopdf}
\usepackage{tabularx}
\usepackage{enumitem}
\DeclareMathOperator{\tr}{tr}

\allowdisplaybreaks
\newcommand\norm[1]{\left\lVert#1\right\rVert}
\DeclareMathOperator{\Gcap}{\widehat{\textbf{G}}}
\DeclareMathOperator{\hcap}{\widehat{\textbf{h}}}
\newcommand\lmin[1]{\lambda_{\text{min}}\left(#1\right)}
\DeclareMathOperator{\Gtilde}{\widetilde{G}}
\DeclareMathOperator{\Htilde}{\widetilde{H}}
\newcommand\tvec[1]{\text{vec}\left(#1\right)}

\begin{document}

\title{A Theoretical Performance Bound for Joint \\ Beamformer Design of Wireless Fronthaul and \\ Access Links in Downlink C-RAN}


\author{Fehmi Emre Kadan,~\IEEEmembership{Student Member, IEEE}, and Ali \"{O}zg\"{u}r Y{\i}lmaz,~\IEEEmembership{Member, IEEE}}


\maketitle

\begin{abstract}
It is known that data rates in standard cellular networks are limited due to inter-cell interference. An effective solution of this problem is to use the multi-cell cooperation idea. In Cloud Radio Access Network (C-RAN), which is a candidate solution in 5G and future communication networks, cooperation is applied by means of central processors (CPs) connected to simple remote radio heads with finite capacity fronthaul links. In this study, we consider a downlink C-RAN with a wireless fronthaul and aim to minimize total power spent by jointly designing beamformers for fronthaul and access links. We consider the case where perfect channel state information is not available in the CP. We first derive a novel theoretical performance bound for the problem defined. Then we propose four algorithms with different complexities to show the tightness of the bound. The first two algorithms apply successive convex optimizations with semi-definite relaxation idea where other two are adapted from well-known beamforming design methods. The detailed simulations under realistic channel conditions show that as the complexity of the algorithm increases, the corresponding performance becomes closer to the bound. 
\end{abstract}


\IEEEpeerreviewmaketitle

\vspace{-5mm}
\section{Introduction}
In new generation communication systems, the number of devices participating in the network grows exponentially. Furthermore, data rate requirements become challenging to satisfy as the network density increases. Standard cellular systems where a set of mobile stations (MSs) are served by a single central base station (BS) have a limited performance due to inter/intra-cell interference. In 5G, Cloud Radio Access Network (C-RAN) is a candidate solution which uses the multi-cell cooperation idea. In C-RAN hierarchy, base stations are simple radio units called remote radio heads (RRHs) which only implement radio functionality such as RF conversions, filtering, and amplifying. All baseband processing is done over a pool of central processors (CPs) which are connected to RRHs with finite capacity fronthaul links. This approach decreases the cost of deployment as compared to the traditional systems where each BS has its own on-site baseband processor. Furthermore, multi-cell cooperation enables better resource allocation and enhances the performance. The main architecture of a typical C-RAN system is described in \cite{C-RAN}.

In a C-RAN cluster of RRHs and MSs, all RRH-to-MS transmissions are performed at the same time and frequency band to use the spectrum efficiently. In traditional C-RAN networks, all RRHs are connected to a CP by means of wired fronthaul links with high capacity. User data is shared among RRHs using fronthaul links enabling an optimized resource allocation. On the other hand, in some situations, the cost of using wired links can be high especially for urban areas. As an alternative approach, one can use a large base station located close to CP to send the user data from CP to RRHs through wireless links. By this method, the rate of data transmission in fronthaul links can be adaptively adjusted using proper power allocations and beamforming schemes. In the wireless fronthaul case, frequency bands of the fronthaul and the access links (links between RRHs and MSs) may be the same or different. In in-band scenario where the two frequency bands are the same, the RRHs should be capable of performing self-interference cancellation which increases the equipment complexity. To make the RRHs simpler, either the two frequency bands may be separated or a time-division based transmission can be used.

In a C-RAN system with wireless fronthaul, the main aim is to design proper beamformers to optimize the network. This problem is similar to a two-hop relay design problem. In relay systems, there are different types of multi-hop mechanisms such as amplify-and-forward (AF), decode-and-forward (DF), decompress-and-forward (DCF), etc. The corresponding method is determined by the operation applied by RRHs to the signal received from fronthaul links before transmitting to users. AF type systems are the simplest ones where RRHs only apply some scaling to the received data \cite{AF-C-RAN}-\cite{AF-Relay2}. In DF based systems, RRHs apply a decoding to the user data requiring baseband processing ability for RRHs \cite{DF-1}-\cite{DF-2}. In DCF based systems, both decoding and decompressing abilities are necessary \cite{DCF-1}-\cite{DCF-2}. In DF and DCF based systems, there is some cooperation between CP and RRHs to decide which RRHs to decode which user data. In general, this requires a combinatorial search making the design complex. On the other hand, as the user data is decoded, assuming a perfect decoding for sufficiently high signal-to-noise-ratio (SNR), the interference between user signals can be eliminated at RRHs allowing to satisfy a higher performance for users. In general, AF systems are simpler but the interference cannot be perfectly eliminated at RRHs. In C-RAN systems, it is intended to make RRHs as simple as possible to decrease the deployment cost making AF systems more attractive.

To optimize a C-RAN network by designing beamformers, channel coefficients should be known with some accuracy. In general, perfect channel state information (CSI) is not available as the channel estimation is done via pilot signals with finite power. There are different models for channel estimation error. It can be shown that linear channel estimation methods with orthogonal pilot signals yield an additive channel estimation error. The error is a random vector whose statistics may be known or not known. Some works assume that first or second order statistics are known \cite{AF-Relay1}, \cite{Ch-Add-1}-\cite{AF-Relay3}, and some other works use the model where error is norm-bounded \cite{AF-C-RAN}, \cite{AF-Relay2}, \cite{AF-Relay4}. The first approach is used when quantization error in channel estimation is negligible and the second one is used when quantization error is dominant \cite{AF-Relay5-ch-err}. Using the knowledge about the channel error vectors, the beamforming design problem can be well optimized and robustness against errors can be achieved. 

In this paper, we consider a downlink C-RAN system with wireless fronthaul where the transmissions of fronthaul and access links are in the same frequency band but in different time-slots. We assume that there is a partial channel knowledge where the second order statistics of the channel error is perfectly known. We optimize fronthaul and access link beamformers with AF type relaying in RRHs. Optimization is performed to minimize total power spent under user signal-to-interference-and-noise-ratio (SINR) constraints. In the literature, the power minimization problem is referred as Quality-of-Service (QoS) \cite{AF-Relay2}. In this approach, it is guaranteed to satisfy a certain quality of service to each user and the total power spent, which is one of the major costs of an operator, is minimized. In this work, our main aim is to find a theoretical lower bound for total power spent in the system. In showing the tightness of a lower bound, existence of an algorithm that comes close to the bound is sufficient since no algorithm can perform better than a lower bound. To show that the given bound is tight enough, we consider four different design methods with different complexities. The first method is Alternating Optimization (AO), which consecutively solves a series of beamforming design problems using convex optimization with semi-definite relaxation (SDR) approach. Both fronthaul and access link beamformers are designed using convex optimizations. The performance of this method is close to the bound but its complexity is high in general. The second method is a modified version of AO which is called Total SNR Maximization (TSM), where fronthaul beamformer design is based on the maximization of total SNR at RRHs. The access link beamformers are found as in AO. The third and fourth methods are proposed as a mixture of standard beamforming design methods which are maximal ratio combining (MRC), zero forcing (ZF) and singular value decomposition (SVD). The third method is a combination of MRC and ZF so it is named as MRC-ZF. In this method, CP beamformers (related to fronthaul link) are found using MRC whereas RRH beamformers (related to access link) are found using ZF. The fourth method is called SVD-ZF and the corresponding CP and RRH beamformers are designed accordingly. MRC-ZF and SVD-ZF can directly find beamformers without using a convex optimization and hence they are simpler compared to AO and TSM. They are considered to make a comparison between the well-known beamforming methods and the high-complexity convex optimization based methods. 

The contributions of the paper can be listed as below: 
\begin{itemize}
\item We derive a theoretical lower bound for total power spent in the system to serve multiple users for a given set of network parameters. By detailed simulations, we show the tightness of the bound. In general, the papers related to C-RAN proposes different design methods whose optimality are not known due to the lack of a theoretical bound or a globally optimum solution. To the best of our knowledge, there is no other work deriving a bound.  
\item We propose four novel design methods. Two of them are based on convex optimization with SDR approach and the other two are based on a combination of well-known methods. Because of the mixed structure of CP and RRH beamformers in SINR expressions, convex optimization cannot be directly applied. We organize the related expressions for which SDR approach is applicable. By similar reasons, the direct application of well-known methods is also not possible. We solve a system of matrix equations to apply MRC, ZF and SVD. 
\item We perform detailed simulations to observe the performances of the proposed methods. We make a comparison to the theoretical bound for different network parameters.
\end{itemize}

The organization of the paper is as follows. In Section II, related works are reviewed. Section III describes the general system model. In Section IV, a novel theoretical performance bound for the proposed problem is derived. Section V includes the convex optimization based methods AO and TSM. The modified beamforming methods MRC-ZF and SVD-ZF are described in Section VI. In Section VII, simulation results are presented. Finally, Section VIII concludes the paper.  

\subsection*{Notation}
Throughout the paper, the vectors are denoted by bold lowercase letters and matrices are denoted by bold uppercase letters. $(\cdot)^T, (\cdot)^H,$ and $\tr(\cdot)$ indicates the transpose, conjugate transpose and trace operators, respectively. $\textbf{0}$ describes the all-zero matrix, and $\textbf{A} \succeq 0$ implies that the matrix $\textbf{A}$ is Hermitian and positive semi-definite. $\text{diag}(x_1, x_2, \ldots, x_n)$ denotes the diagonal matrix with diagonal elements $x_1, x_2, \ldots, x_n$ and $\textbf{I}_n$ denotes $n \times n$ identity matrix. $\lmin{\cdot}, \lambda_i(\cdot), e_i(\cdot)$ denotes the minimum eigenvalue, $i$-th largest eigenvalue and the corresponding unit-norm eigenvector of the corresponding Hermitian positive semi-definite matrix, respectively. $\norm{\cdot}$ denotes the $\ell_2$-norm of the corresponding matrix, $\mathbb{E}[\cdot]$ denotes the expectation operator. $\tvec{\textbf{A}}$ is the column vector consisting of columns of $\textbf{A}$. The symbol $\otimes$ denotes the Kronecker product. Finally, $\mathbb{C}$ denotes the set of complex numbers and $\delta[\cdot]$ corresponds to the discrete impulse function satisfying $\delta[0]=1, \: \delta[x]=0$ for all $x \neq 0$.

\section{Related Studies}

In this section, we review related studies existing in the literature. Firstly, we present the works related to wired fronthaul links and mention the main differences compared to the wireless case. Secondly, we review the studies related to AF, DF and DCF type wireless fronthaul systems and indicate the main differences with our work. Thirdly, we mention papers with different channel uncertainty models used in C-RAN system designs. Finally, we express the major differences of the papers related to standard relay networks. 

\vspace{-5mm}

\subsection{Wired Fronthaul}

There are a lot of studies existing in the literature related to multi-cell cooperation techniques for wired fronthaul. In \cite{Wired-Rate1}-\cite{Wired-Rate3}, optimization is performed to maximize the data rate of users under certain transmit power and fronthaul capacity constraints. The optimization of SINRs of users is analyzed in \cite{Wired-UDD} using uplink-downlink duality. In \cite{Wired-LimitedFronthaul}, the total transmit power is minimized under fronthaul capacity constraints. In \cite{Wired-SDR}, the cost function consists of a weighted sum of the total transmit power and the total fronthaul data. As another approach, \cite{Wired-UserMax} aims at finding the largest set of users which can be served by the system where each user data is sent only by a single RRH. The power consumption of RRHs under active and sleeping modes can also be included to the power minimization problem as done in \cite{Wired-GreenCRAN}. In \cite{Wired-ZF}, a standard ZF beamformer design is used, however its performance is limited in eliminating the interference. In \cite{Wired-Heuristic1}-\cite{Wired-Heuristic3}, the cooperation strategy is found using some heuristic search techniques. Possible strategies in imperfect channel case are considered in \cite{Ch-Add-3}, \cite{Wired-Imperfect-CSI}. Cluster formation \cite{Wired-ClusterFormation} and the effect of user traffic delay \cite{Wired-Delay} are also analyzed in the literature. 

For wired fronthaul case, as there is no interference between different users at RRHs, there is a natural combinatorial user selection problem. CP determines the set of users to be served by each RRH (possibly intersecting) and sends the corresponding data through fronthaul links. In general, most of the studies assume that perfect user data is available at RRHs after fronthaul transmission where some works also take the decompression error effect into account. Since the fronthaul transmission takes places over cables, there is no beamforming in CP. The design problem is to decide on the cooperation strategy and beamforming coefficients for access link. On the other hand, in wireless fronthaul networks, both fronthaul and access links have their own beamformers which are the main design parameters. Considering the differences in fronthaul structures, the methods proposed for wired case cannot be directly applied to wireless case. 

\vspace{-4mm}

\subsection{Relaying Mechanism for Wireless Fronthaul}

Works related to wireless fronthaul case are limited in number compared to the standard wired case. The problem for wireless fronthaul case is similar to two-hop relaying. Most studied relaying mechanisms for C-RAN with wireless fronthaul concept are AF, DF and DCF. In \cite{DF-1} DF based relaying is assumed where each RRH can decode only a single user's data at once. If more than one user's data is to be decoded, decoding is done by time division. The combinatorial problem of choosing the set of user data to be decoded by each RRH is solved in \cite{DF-1} while an SDR based beamformer optimization is done under perfect CSI assumption. \cite{DF-2} also analyzes DF based relaying where a weighted sum of user data rates is maximized under power transmit limit. There is a constraint that each RRH can serve a single user. Beamformer optimization is performed using SDR and perfect CSI is assumed. In \cite{DCF-1} both DF and DCF based approaches are considered where the set of user data to be decoded by each RRH is assumed to be known and beamforming optimization is done using difference of convex method. Data rate maximization under power limit is analyzed for perfect CSI case. \cite{DCF-2} is the generalized version of \cite{DCF-1} where there are more than one RRH clusters each controlled by a different CP. \cite{AF-C-RAN} uses AF type relaying with a norm-bounded channel estimation error model. Using worst-case SINR formulas total power is minimized under SINR constraints. In that work, fronthaul beamformers are assumed to be known and access link beamformers are designed using SDR based methods along with a ZF based approach implemented for comparison. In \cite{AF-Relay2}, a two-hop AF relaying problem is studied under norm-bounded channel error model. As all independent sources have a single antenna, fronthaul beamforming is not applicable, only access link beamforming design is studied. SDR based optimization is used to minimize total transmit power under SINR constraints.

Because of the combinatorial nature of DF and DCF based relaying schemes, the methods used for fronthaul beamforming design cannot be directly adapted to AF type relaying. For access link beamforming design, SDR based approach is widely used for all types of relaying schemes. Some works also consider well known beamforming methods (such as ZF) for comparison. To the best of our knowledge, there is a very limited amount of work about C-RAN with wireless fronthaul and AF relaying. Furthermore, in such studies, neither the fronthaul and access link beamforming design is jointly considered nor a theoretical bound is derived. 

\vspace{-5mm}

\subsection{Channel Error Model}

In C-RAN concept, three types of channel error models are mostly used. The first one is perfect CSI model where channel coefficients are assumed to be perfectly known. Although it is unrealistic, the methods proposed for this case may provide some insights. Furthermore, in most of the cases, it is possible to modify the corresponding algorithms accordingly when the channel is partially known. The papers \cite{DF-1}-\cite{DCF-2}, \cite{Wired-Rate1}-\cite{Wired-Heuristic3} all assume perfect CSI. The second approach is the norm-bounded error assumption. In this assumption, it is assumed that the error vectors are in some sphere with known radius. The works with this assumption perform beamforming design using worst case SINRs which can be defined as the minimum value of SINRs for given error norm bounds. \cite{AF-C-RAN}, \cite{AF-Relay2}, \cite{AF-Relay4}-\cite{AF-Relay5-ch-err} and some references therein use this method. The third approach which is also used in our work assumes that second order statistics (mean and covariance matrices) of the channel estimation error vectors are known. When this approach is used the mean powers of signal, interference and noise terms are used in the design process. \cite{AF-Relay1}, \cite{Ch-Add-1}-\cite{AF-Relay3} use the last approach. 

\vspace{-5mm}

\subsection{Standard Relay Networks}

The C-RAN with wireless fronthaul concept is similar to two-hop multi source/destination multi-antenna relaying networks and some beamforming design techniques used in standard relaying literature (such as SDR) can be adapted to C-RAN framework. On the other hand, joint optimization of fronthaul and access link beamformers is not widely considered in standard relaying problems. \cite{AF-Relay1}-\cite{AF-Relay2}, \cite{AF-Relay3}-\cite{AF-Relay5-ch-err}, \cite{AF-Relay6}-\cite{AF-Relay7} include beamforming design for standard relaying problems which are all special cases of our problem of concern. Hence, some methods proposed for relaying problems can be used for our purposes but none of them directly provides a solution. 

\section{System Model}

We consider the downlink of a C-RAN cluster including a CP with $M$ antennas, $N$ RRHs each with $L$ transmit/receive antennas, and $K$ MSs each with a single antenna. All CP-to-RRH and RRH-to-MS channels are assumed to be flat, constant over a transmission period and known by CP with some additive Gaussian error with known second order statistics. We assume a two stage transmission scheme where fronthaul and access link transmissions are performed in different time slots. In the first stage, the user data is sent from CP to RRHs over wireless channels. RRHs apply some linear transformation to the received data as in AF relaying using beamforming matrices and forward the transformed signal to the MSs in the second stage. We assume that RRHs are simple radio units without the capability of baseband processing and hence they cannot decode the user data. Therefore, AF relaying mechanism is considered in this model. In Fig. 1, we see the general block diagram of the model used. 

\vspace{-5mm}

\begin{figure}[H]
\centering
\captionsetup{justification=centering}
\includegraphics[width=0.62\textwidth]{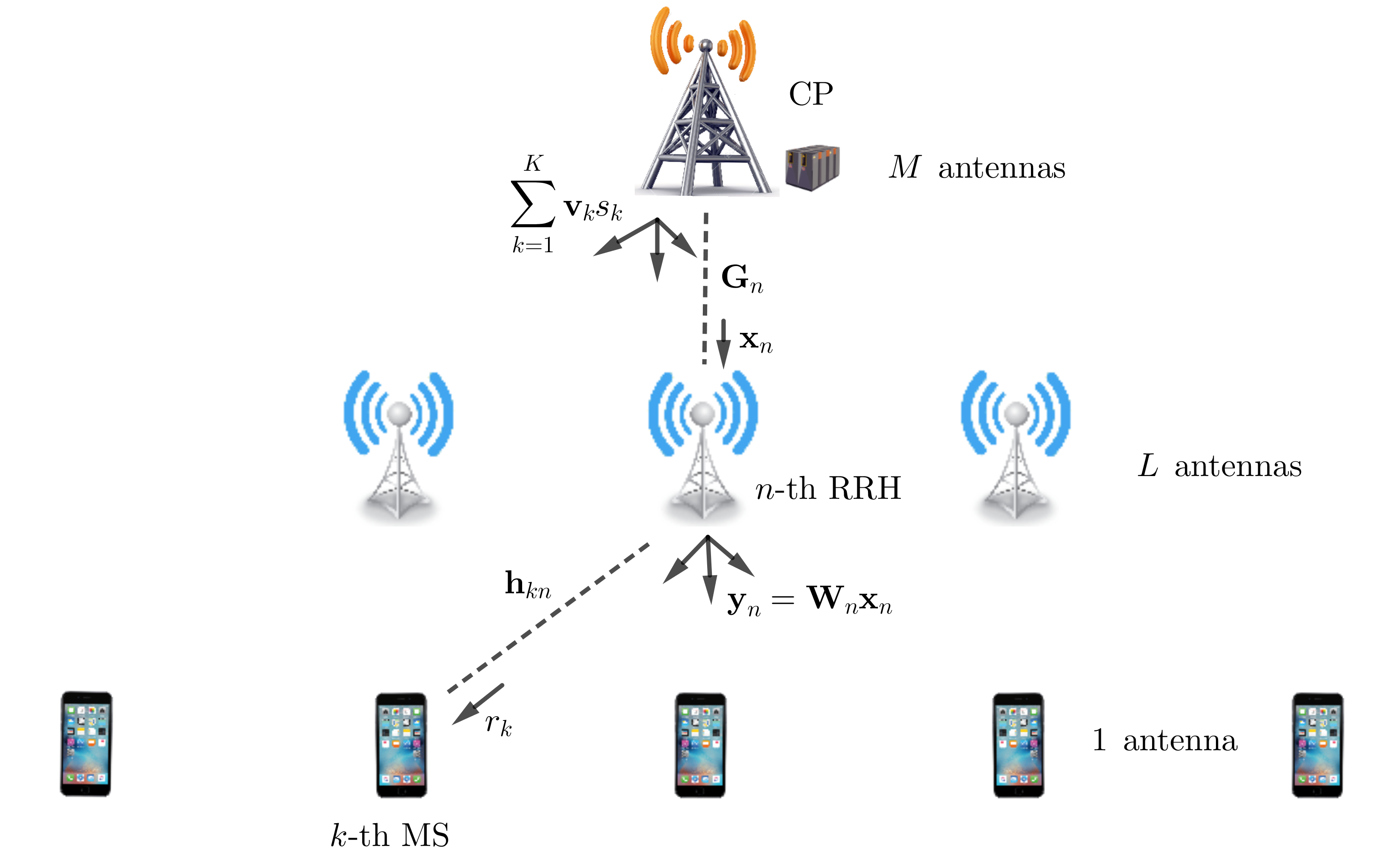}
\caption{Block Diagram of Downlink C-RAN with Wireless Fronthaul.}
\end{figure}

\vspace{-5mm}

We denote the channel between CP and $n$-th RRH as $\textbf{G}_n \in \mathbb{C}^{M \times L}$, the channel between $n$-th RRH and $k$-th MS as $\textbf{h}_{kn} \in \mathbb{C}^L$, the beamformer vector of CP for $k$-th user as $\textbf{v}_{k}\in \mathbb{C}^M$, and beamforming matrix for $n$-th RRH as $\textbf{W}_n\in \mathbb{C}^{L \times L}$. The received signal of the $n$-th RRH in the first transmission stage can be written as
\begin{equation}
\textbf{x}_n = \textbf{G}_n^H\displaystyle\sum_{k=1}^K\textbf{v}_{k}s_k+\textbf{z}_n, \quad n=1, 2, \ldots, N
\end{equation} 
where $s_k$ denotes the $k$-th user data which satisfies $\mathbb{E}[|s_k|^2]=1, \quad \forall k=1, 2, \ldots, K$ and $\textbf{z}_n \sim \mathcal{C}\mathcal{N}(\textbf{0}, \sigma_{\text{RRH}}^2\textbf{I}_{L})$ is the noise term in the corresponding RRH. After the first stage, the transformed signal by $n$-th RRH is given by 
\begin{equation}
\textbf{y}_n = \textbf{W}_n \textbf{x}_n, \quad n=1, 2, \ldots, N.
\end{equation} 
In this case, the received signal by the $k$-th MS can be expressed by
\begin{equation}
\begin{aligned}
r_k &= \displaystyle\sum_{n=1}^N \textbf{h}_{kn}^H\textbf{y}_n + n_k = \displaystyle\sum_{n=1}^N \textbf{h}_{kn}^H \textbf{W}_n \left(\textbf{G}_n^H\displaystyle\sum_{\ell=1}^K\textbf{v}_{\ell}s_{\ell} + \textbf{z}_n\right) + n_k \\
&= \displaystyle\sum_{n=1}^N \textbf{h}_{kn}^H \textbf{W}_n \textbf{G}_n^H \textbf{v}_{k}s_{k}+\displaystyle\sum_{n=1}^N \displaystyle\sum_{\ell \neq k}^K \textbf{h}_{kn}^H \textbf{W}_n \textbf{G}_n^H \textbf{v}_{\ell}s_{\ell} + \displaystyle\sum_{n=1}^N \textbf{h}_{kn}^H \textbf{W}_n \textbf{z}_n + n_k. 
\end{aligned}
\end{equation}
Here $n_k \sim \mathcal{C}\mathcal{N}(0, \sigma_{\text{MS}}^2)$ denotes the noise term in the $k$-th MS. In order to simplify expressions, we define augmented channel, beamformer and noise vectors/matrices as given below:
\begin{equation}
\begin{aligned}
\textbf{h}_k &= [\textbf{h}_{k1}^T \: \textbf{h}_{k2}^T \: \cdots \: \textbf{h}_{kN}^T]^T \: : \: NL \times 1, \quad \textbf{W} = \text{diag}\left(\textbf{W}_1, \: \textbf{W}_2, \: \ldots, \: \textbf{W}_N \right) \: : \: NL \times NL, \\
\textbf{G} &= \left[\textbf{G}_1 \: \textbf{G}_2 \: \cdots \: \textbf{G}_N \right] \: : \: M \times NL, \quad 
\textbf{z} = [ \textbf{z}_1^T \: \textbf{z}_2^T \: \cdots \: \textbf{z}_N^T]^T \: : \: NL \times 1. \\
\end{aligned}
\end{equation}
Using the augmented variables, we can write $r_k$ as 
\begin{equation} \label{r_k_1}
r_k = \textbf{h}_k^H\textbf{W}\textbf{G}^H\textbf{v}_k s_k + \displaystyle\sum_{\ell\neq k} \textbf{h}_k^H\textbf{W}\textbf{G}^H\textbf{v}_{\ell}s_{\ell} + \textbf{h}_k^H\textbf{W}\textbf{z} + n_k. 
\end{equation}
We model the channel estimates as $\textbf{G}_n = \widehat{\textbf{G}}_n + \Delta \textbf{G}_n, \: \: \textbf{h}_{kn} = \widehat{\textbf{h}}_{kn} + \Delta \textbf{h}_{kn}$ where $\widehat{\textbf{G}}_n $ and $\widehat{\textbf{h}}_{kn}$ are channel estimates, $\Delta \textbf{G}_n$ is a zero-mean complex Gaussian matrix with independent entries each with variance $\sigma_{1,n}^2$ and $\Delta \textbf{h}_{kn} \sim \mathcal{C}\mathcal{N}\left(\textbf{0}, \sigma_{2,k,n}^2\textbf{I}_{L}\right)$ is a circularly symmetric Gaussian vector. We also assume that $\Delta \textbf{G}_n$ and $\Delta \textbf{h}_{kn}$ are independent for all $n$ and $k$. Using the error vectors and matrices, we can form the corresponding augmented variables as shown in (\ref{h_aug}). 
\begin{equation} \label{h_aug}
\begin{aligned}
\widehat{\textbf{h}}_k &= [\widehat{\textbf{h}}_{k1}^T \: \widehat{\textbf{h}}_{k2}^T \: \cdots \: \widehat{\textbf{h}}_{kN}^T]^T \: : \: NL \times 1, \quad 
\widehat{\textbf{G}} = \left[\widehat{\textbf{G}}_1 \: \widehat{\textbf{G}}_2 \: \cdots \: \widehat{\textbf{G}}_N \right] \: : \: M \times NL, \\
\Delta \textbf{h}_k &= [\Delta \textbf{h}_{k1}^T \: \Delta \textbf{h}_{k2}^T \: \cdots \: \Delta \textbf{h}_{kN}^T]^T \: : \: NL \times 1, \quad 
\Delta \textbf{G} = \left[\Delta \textbf{G}_1 \: \Delta \textbf{G}_2 \: \cdots \: \Delta \textbf{G}_N \right] \: : \: M \times NL.
\end{aligned}  
\end{equation}
Using the new variables and (\ref{r_k_1}), we can write $r_k$ as
\begin{equation} \label{r_k_2}
r_k = \underbrace{\widehat{\textbf{h}}_k^H\textbf{W}\widehat{\textbf{G}}^H\textbf{v}_k s_k}_{\text{desired}} + \underbrace{\left(\textbf{h}_k^H\textbf{W}\textbf{G}^H\textbf{v}_k - \widehat{\textbf{h}}_k^H\textbf{W}\widehat{\textbf{G}}^H\textbf{v}_k\right) s_k}_{\text{interference part 1}} + \underbrace{\displaystyle\sum_{\ell\neq k} \textbf{h}_k^H\textbf{W}\textbf{G}^H\textbf{v}_{\ell}s_{\ell}}_{\text{interference part 2}} + \underbrace{\textbf{h}_k^H\textbf{W}\textbf{z} + n_k}_{\text{noise}}.
\end{equation}
In (\ref{r_k_2}), the desired part includes the desired signal for the $k$-th MS. Notice that it contains only the channel estimates for the $k$-th user which is the only useful part for the receiver of corresponding MS. Interference part 1 is related to the channel mismatch for the $k$-th user signal. Although it includes $s_k$ term, the corresponding signal is not useful as its coefficient is not known by the receiver due to uncertainty in the channel estimates. Interference part 2 is the actual interference signal including the signals for other users. Noise term is the combination of the amplified and forwarded RRH receiver noise and MS receiver noise. Using the equation in (\ref{r_k_2}), we define
\begin{equation} \label{SINR_1}
\text{SINR}_k = \dfrac{P_d}{P_{I,1}+P_{I,2}+P_n} 
\end{equation}
where
\begin{equation} \label{P_d}
\begin{aligned}
P_d &= \mathbb{E}\left\{\left|\widehat{\textbf{h}}_k^H\textbf{W}\widehat{\textbf{G}}^H\textbf{v}_k s_k\right|^2\right\}, \: \:
&&P_{I,1} = \mathbb{E}\left\{\left|\left(\textbf{h}_k^H\textbf{W}\textbf{G}^H\textbf{v}_k - \widehat{\textbf{h}}_k^H\textbf{W}\widehat{\textbf{G}}^H\textbf{v}_k\right) s_k \right|^2\right\} \\
P_{I,2} &= \mathbb{E}\left\{\left|\displaystyle\sum_{\ell\neq k} \textbf{h}_k^H\textbf{W}\textbf{G}^H\textbf{v}_{\ell}s_{\ell} \right|^2\right\}, \: \:
&&P_n = \mathbb{E}\left\{\left|\textbf{h}_k^H\textbf{W}\textbf{z} + n_k \right|^2\right\}.
\end{aligned}
\end{equation}
Using the fact that $\mathbb{E}\left[s_k^Hs_{\ell}\right]=\delta[k-\ell]$ and statistics of the channel error matrices/vectors and noise terms, we find that 
\begin{equation} \label{SINR_2}
\text{SINR}_k = \dfrac{\left|\widehat{\textbf{h}}_k^H\textbf{W}\widehat{\textbf{G}}^H\textbf{v}_k\right|^2}{\displaystyle\sum_{\ell=1}^K\tr\left(\textbf{D}_k\textbf{W}\textbf{C}_{\ell}\textbf{W}^H\right) - \left|\widehat{\textbf{h}}_k^H\textbf{W}\widehat{\textbf{G}}^H\textbf{v}_k\right|^2 + \sigma_{\text{RRH}}^2\tr\left(\textbf{D}_k\textbf{W}\textbf{W}^H\right) + \sigma_{\text{MS}}^2}
\end{equation}
where
\begin{equation}
\begin{aligned}
\textbf{D}_k &= \widehat{\textbf{h}}_k\widehat{\textbf{h}}_k^H + \bm{\Sigma}_{2,k}, \: \: 
\textbf{C}_k = \widehat{\textbf{G}}^H\textbf{v}_k\textbf{v}_k^H \widehat{\textbf{G}} + (\textbf{v}_k^H\textbf{v}_k)\bm{\Sigma}_1, \\
\bm{\Sigma}_1 &= \text{diag}\left(\sigma_{1,1}^2\textbf{I}_L, \sigma_{1,2}^2\textbf{I}_L, \ldots, \sigma_{1,N}^2\textbf{I}_L\right), \: \:
\bm{\Sigma}_{2,k} = \text{diag}\left(\sigma_{2,k,1}^2\textbf{I}_L, \sigma_{2,k,2}^2\textbf{I}_L, \ldots, \sigma_{2,k,N}^2\textbf{I}_L\right).
\end{aligned}
\end{equation}
In Appendix A, we show that the rate $\log_2(1+\text{SINR}_k)$ is achievable for the $k$-th user. Hence, the SINR that we defined can be used as a design criteria. Another design term that can be optimized is the total power spent in the system. The total power $P$ has two components $P_{\text{CP}}$ and $P_{\text{RRH}}$ which correspond to the power transmitted by CP and RRHs, respectively. Using the fact that $\mathbb{E}\left[s_k^H s_{\ell}\right]=\delta[k-\ell]$, we can write\footnote{Actual power terms include a constant multiplier which does not affect the solution, and hence they are omitted.}
\begin{equation}
P_{\text{CP}} = \mathbb{E}\left[\left|\displaystyle\sum_{k=1}^K\textbf{v}_ks_k\right|^2\right] \\
= \displaystyle\sum_{k=1}^K \textbf{v}_k^H \textbf{v}_k,
\end{equation}
and
\begin{equation}
\begin{aligned}
P_{\text{RRH}} &= \displaystyle\sum_{n=1}^N \mathbb{E}\left[\left|\textbf{y}_n\right|^2\right] = \displaystyle\sum_{n=1}^N \mathbb{E}\left[\left|\textbf{W}_n\left(\textbf{G}_n^H\displaystyle\sum_{k=1}^K\textbf{v}_{k}s_k+\textbf{z}_n\right)\right|^2\right] \\
&= \displaystyle\sum_{k=1}^K \textbf{v}_k^H \textbf{G}\textbf{W}^H \textbf{W} \textbf{G}^H \textbf{v}_k + \sigma_{\text{RRH}}^2 \tr\left(\textbf{W}^H \textbf{W}\right).
\end{aligned}
\end{equation}
Due to imperfect channel state information, $P_{\text{RRH}}$ includes random terms. Therefore, we optimize the mean power $P=P_{\text{CP}} + \mathbb{E}\left\{P_{\text{RRH}}\right\}$ which can be evaluated as 
\begin{equation} \label{P_eqn}
P = \displaystyle\sum_{k=1}^K \tr\left(\bm{\tau_0} \textbf{v}_k\textbf{v}_k^H\right) + \sigma_{\text{RRH}}^2 \tr\left(\textbf{W}^H \textbf{W}\right)
\end{equation}
where $\bm{\tau_0}= \textbf{I}_M + \widehat{\textbf{G}}\textbf{W}^H\textbf{W}\widehat{\textbf{G}}^H + \tr\left(\textbf{W}^H \textbf{W} \bm{\Sigma}_1\right)\textbf{I}_M$.
In this study, we aim to minimize total mean power $P$ under SINR constraints $\text{SINR}_k \geq \gamma_k$ where $\left\{\gamma_k\right\}_{k=1}^K$ are given SINR thresholds.\footnote{Feasibility cannot be guaranteed. Bad channel conditions and/or high SINR thresholds may yield infeasible results.} As shown in Appendix A, the SINR constraints provide that the rate $\log_2(1+\gamma_k)$ is achievable for the $k$-th user. This type of problem is studied under Quality-of-Service (QoS) in the literature where we minimize the power spent in the system by satisfying a certain rate (or SINR) for each user. User rates can be adjusted according to the priority of users by changing the corresponding threshold values. The main optimization problem (P0) can be formulated as 
\begin{equation}
(\text{P}0) \: \: \min_{\textbf{W}, \{\textbf{v}_k\}_{k=1}^K} P \quad \text{such that} \quad \text{SINR}_k \geq \gamma_k, \quad \forall k=1, 2, \ldots, K.
\end{equation}

\section{A Theoretical Performance Bound}
In this section, we find a novel performance bound for (P0). We find a lower bound for the total mean power P under SINR constraints. Using the SINR constraints, for all $k$ we have
\begin{equation} \label{B1}
\left|\widehat{\textbf{h}}_k^H\textbf{W}\widehat{\textbf{G}}^H\textbf{v}_k\right|^2 \geq \gamma_k\left(\tr\left(\textbf{D}_k\textbf{W}\textbf{C}_k\textbf{W}^H\right) - \left|\widehat{\textbf{h}}_k^H\textbf{W}\widehat{\textbf{G}}^H\textbf{v}_k\right|^2 + \sigma_{\text{RRH}}^2\tr\left(\textbf{D}_k\textbf{W}\textbf{W}^H\right) + \sigma_{\text{MS}}^2\right).
\end{equation}
Numerical manipulations reveal that 
\begin{equation} \label{B2}
\begin{aligned}
\tr\left(\textbf{D}_k\textbf{W}\textbf{C}_k\textbf{W}^H\right) - \left|\widehat{\textbf{h}}_k^H\textbf{W}\widehat{\textbf{G}}^H\textbf{v}_k\right|^2 &= \tr\left(\bm{\Sigma}_{2,k} (\textbf{W}\Gcap^H\textbf{v}_k)(\textbf{W}\Gcap^H\textbf{v}_k)^H\right) + \\ &\textbf{v}_k^H \textbf{v}_k\left[\tr\left((\hcap_k^H\textbf{W})^H(\hcap_k^H\textbf{W})\bm{\Sigma}_1\right) + \tr\left(\textbf{W}^H\textbf{W}\bm{\Sigma}_{2,k}\bm{\Sigma}_1\right)\right].
\end{aligned}
\end{equation}
To show (\ref{B2}), we use the facts $\textbf{W}\bm{\Sigma}_1=\bm{\Sigma}_1\textbf{W}$ and $\textbf{W}\bm{\Sigma}_{2,k}=\bm{\Sigma}_{2,k}\textbf{W}$. We know that using Von-Neumann's Inequality \cite{Von-Neumann}, for any two $c \times c$ Hermitian positive semi-definite matrices $\textbf{A}$ and $\textbf{B}$ we have $\tr\left(\textbf{A}\textbf{B}\right) \geq \displaystyle\sum_{i=1}^c \lambda_i(\textbf{A}) \lambda_{c-i+1}(\textbf{B}) \geq \lambda_c(\textbf{A}) \lambda_1(\textbf{B}) = \lmin{\textbf{A}}\norm{\textbf{B}}$. Using this fact and (\ref{B2}), we have
\begin{equation}
\begin{aligned}
\tr\left(\textbf{D}_k\textbf{W}\textbf{C}_k\textbf{W}^H\right) - \left|\widehat{\textbf{h}}_k^H\textbf{W}\widehat{\textbf{G}}^H\textbf{v}_k\right|^2 &\geq \lmin{\bm{\Sigma}_{2,k}}\norm{\textbf{W}\Gcap^H\textbf{v}_k}^2 + \\
&\textbf{v}_k^H\textbf{v}_k\left[\lmin{\bm{\Sigma}_1}\norm{\hcap_k^H\textbf{W}}^2 + \lmin{\bm{\Sigma}_{2,k}\bm{\Sigma}_1}\norm{\textbf{W}}^2\right].
\end{aligned}
\end{equation}
Similarly, we get $\tr\left(\textbf{D}_k\textbf{W}\textbf{W}^H\right) = \tr\left((\hcap_k\hcap_k^H+\bm{\Sigma}_{2,k})\textbf{W}\textbf{W}^H\right) \geq \norm{\hcap_k\textbf{W}}^2+\lmin{\bm{\Sigma}_{2,k}}\norm{\textbf{W}}^2$. Therefore, we obtain that
\begin{align} \label{B3}
\left|\widehat{\textbf{h}}_k^H\textbf{W}\widehat{\textbf{G}}^H\textbf{v}_k\right|^2 &\geq \gamma_k \Big[\lmin{\bm{\Sigma}_{2,k}}\norm{\textbf{W}\Gcap^H\textbf{v}_k}^2 + \lmin{\textbf{v}_k^H\textbf{v}_k\bm{\Sigma}_1+\sigma_{\text{RRH}}^2\textbf{I}_{NL}}\norm{\hcap_k^H\textbf{W}}^2 + \notag \\
&\lmin{\textbf{v}_k^H\textbf{v}_k\bm{\Sigma}_{2,k}\bm{\Sigma}_1 + \sigma_{\text{RRH}}^2\bm{\Sigma}_{2,k}}\norm{\textbf{W}}^2 + \sigma_{\text{MS}}^2 \Big] \\
&= \gamma_k \left[\sigma_{2,k}^2\norm{\textbf{W}\Gcap^H\textbf{v}_k}^2+(\textbf{v}_k^H\textbf{v}_k\sigma_1^2+\sigma_{\text{RRH}}^2)\left(\norm{\hcap_k^H\textbf{W}}^2 + \sigma_{2,k}^2\norm{\textbf{W}}^2\right) + \sigma_{\text{MS}}^2 \right] \notag
\end{align}
where $\sigma_1^2 = \min\limits_{n} \sigma_{1,n}^2$ and $\sigma_{2,k}^2=\min\limits_{n} \sigma_{2,k,n}^2$. Similarly, we obtain that
\begin{equation} \label{B4}
\begin{aligned}
P &\geq \displaystyle\sum_{k=1}^K \left(\textbf{v}_k^H\textbf{v}_k + \norm{\textbf{W}\Gcap^H\textbf{v}_k}^2 +  \textbf{v}_k^H\textbf{v}_k \lmin{\bm{\Sigma}_1}\norm{\textbf{W}}^2\right)+\sigma_{\text{RRH}}^2\tr\left(\textbf{W}^H\textbf{W}\right) \\
&\geq \displaystyle\sum_{k=1}^K \Bigl( \underbrace{\textbf{v}_k^H\textbf{v}_k + \norm{\textbf{W}\Gcap^H\textbf{v}_k}^2 + \textbf{v}_k^H\textbf{v}_k\sigma_1^2\norm{\textbf{W}}^2 + \dfrac{\sigma_{\text{RRH}}^2}{K}\norm{\textbf{W}}^2}_{P_k} \Bigr).
\end{aligned}
\end{equation}
We will find a lower bound for $P_k$ for all $k$ using (\ref{B3}). To simplify the notations, we define 
\begin{equation} \label{B_def}
\begin{aligned}
x_1=\left|\widehat{\textbf{h}}_k^H\textbf{W}\widehat{\textbf{G}}^H\textbf{v}_k\right|^2, \: x_2=\norm{\textbf{W}\Gcap^H\textbf{v}_k}^2, \: x_3&=\norm{\hcap_k^H\textbf{W}}^2, \: x_4=\norm{\textbf{W}}^2, \: x_5=\textbf{v}_k^H\textbf{v}_k, \: y=P_k \\
c_1=\gamma_k, \: c_2=\sigma_{2,k}^2, \: c_3= \sigma_1^2, \: c_4=\sigma_{\text{RRH}}^2, \: c_5&=\sigma_{\text{MS}}^2, \: c_6 = \dfrac{\sigma_{\text{RRH}}^2}{K}, \: d_1=\norm{\hcap_k}^2, \: d_2 = \norm{\Gcap}^2.
\end{aligned}
\end{equation}
(\ref{B3}) and (\ref{B4}) can be written in terms of new variables as 
\begin{equation} \label{B5}
x_1 \geq c_1\left[c_2x_2 + (c_3x_5+c_4)(x_3+c_2x_4)+c_5\right], \: \: 
y = x_2+x_5+c_3x_4x_5+c_6x_4.
\end{equation}
By Cauchy-Schwarz Inequality \cite{CS} and submultiplicativity of $\ell_2$-norm, we get
\begin{equation} \label{B6}
\norm{\hcap_k}^2 \norm{\textbf{W}\Gcap^H\textbf{v}_k}^2 \geq \left|\widehat{\textbf{h}}_k^H\textbf{W}\widehat{\textbf{G}}^H\textbf{v}_k\right|^2 \: \Longrightarrow \: x_2 d_1 \geq x_1. 
\end{equation}
\begin{equation} \label{B7}
\norm{\textbf{W}}^2 \norm{\hcap_k}^2 \geq \norm{\hcap_k^H\textbf{W}}^2 \: \Longrightarrow \: x_4 d_1 \geq x_3. 
\end{equation}
\begin{equation} \label{B8}
\norm{\hcap_k^H\textbf{W}}^2 \norm{\textbf{v}_k}^2 \norm{\Gcap}^2 \geq \norm{\hcap_k^H\textbf{W}}^2 \norm{\textbf{v}_k^H\Gcap}^2 \geq \left|\widehat{\textbf{h}}_k^H\textbf{W}\widehat{\textbf{G}}^H\textbf{v}_k\right|^2 \: \Longrightarrow \: x_3 x_5 d_2 \geq x_1. 
\end{equation}
In Appendix B, using (\ref{B5})-(\ref{B8}) and Arithmetic-Geometric Mean Inequality \cite{AM-GM}, we show that 
\begin{equation} \label{B_bound}
y \geq \dfrac{1}{a}\left(b+c_3c_5+2\sqrt{c_3c_5b+c_5(d_2+c_6)a}\right),
\end{equation}
where $a=\dfrac{d_1d_2}{c_1}-c_2d_2-c_3d_1-c_2c_3, \: b=c_4(d_1+c_2)$. Together with the feasibility condition also found in Appendix B, we can express the bound as 
\begin{equation}
P \geq \displaystyle\sum_{k=1}^K \dfrac{\Htilde_k\sigma_{\text{RRH}}^2+\Gtilde\sigma_{\text{MS}}^2+2\sigma_{\text{RRH}}\sigma_{\text{MS}}\sqrt{\Htilde_k\Gtilde+\dfrac{\Delta_k}{K}}}{\Delta_k}, \: \: \Delta_k>0, \: \forall k
\end{equation}
where
\begin{equation} 
\Htilde_k = \norm{\hcap_k}^2+\sigma_{2,k}^2, \: \Gtilde = \norm{\Gcap}^2+\sigma_1^2, \: \Delta_k=\left(1+\dfrac{1}{\gamma_k}\right)\norm{\hcap_k}^2\norm{\Gcap}^2-\Htilde_k\Gtilde.
\end{equation}
Using (\ref{B_def}), it can be shown that $a=\Delta_k$. In Appendix B, we show that $a>0$ (equivalently $\Delta_k>0, \: \forall k$) is a necessary (but not sufficient) feasibility condition which has to be satisfied to obtain a proper solution for (P0).\footnote{We can find upper bounds for SINR thresholds considering $\Delta_k=0$ to obtain a feasible solution.} It it easy to show that the lower bound is an increasing function of $\sigma_{\text{RRH}}, \sigma_{\text{MS}}, \sigma_1, \sigma_{2,k}, \gamma_k$ and a decreasing function of $\norm{\hcap_k}$ and $\norm{\Gcap}$, as expected. 
 
\section{Convex Optimization Methods}
In the previous section, we have found a performance bound for problem (P0). To observe the tightness of the proposed lower bound, we consider different methods to solve the joint beamformer design problem. In this section, we present two convex optimization based methods to solve (P0). Both methods apply successive convex optimizations with the SDR idea. Firstly, we will show that each one of fronthaul and access link beamformers can be found using convex optimization with SDR when the other one is fixed. Using this observation, we will propose two methods with different complexities. 

\vspace{-5mm}

\subsection{Access Link Beamformer Design}
Let $\textbf{v}_k$'s be given. In this case, the matrices $\textbf{D}_k$ and $\textbf{C}_{\ell}$ become constant. For any matrices $\textbf{X}, \textbf{Y}, \textbf{Z}$ with suitable dimensions, we have $\tr\left(\textbf{X}^H\textbf{Y}\textbf{X}\textbf{Z}\right) = (\tvec{\textbf{X}})^H\left(\textbf{Z}^T\otimes\textbf{Y}\right)\tvec{\textbf{X}}$ \cite{vec-eqn}. Using this property, we get
\begin{align} \label{C1}
\left|\widehat{\textbf{h}}_k^H\textbf{W}\widehat{\textbf{G}}^H\textbf{v}_k\right|^2 &= (\tvec{\textbf{W}})^H\left((\Gcap^H\textbf{v}_k\textbf{v}_k^H\Gcap)^T \otimes (\hcap_k\hcap_k^H) \right) \tvec{\textbf{W}} \notag \\
\tr\left(\textbf{D}_k\textbf{W}\textbf{C}_{\ell}\textbf{W}^H\right) &= (\tvec{\textbf{W}})^H\left(\textbf{C}_{\ell}^T \otimes \textbf{D}_k\right) \tvec{\textbf{W}}, \\
\tr\left(\textbf{D}_k\textbf{W}\textbf{W}^H\right) &= (\tvec{\textbf{W}})^H\left(\textbf{I}_{NL} \otimes \textbf{D}_k\right) \tvec{\textbf{W}}, \: \: \tr\left(\textbf{W}^H\textbf{W}\right) = (\tvec{\textbf{W}})^H\tvec{\textbf{W}}. \notag
\end{align}
Similarly, we obtain that
\begin{align} \label{C2}
\tr\left(\bm{\tau_0} \textbf{v}_k\textbf{v}_k^H\right) &= \textbf{v}_k^H\textbf{v}_k + \tr\left(\textbf{W}^H\textbf{W}\Gcap^H\textbf{v}_k\textbf{v}_k^H\Gcap\right) + \tr\left(\textbf{W}^H\textbf{W}\bm{\Sigma}_1\right)\textbf{v}_k^H\textbf{v}_k \notag \\
&= \textbf{v}_k^H\textbf{v}_k + (\tvec{\textbf{W}})^H\left((\Gcap^H\textbf{v}_k\textbf{v}_k^H\Gcap)^T \otimes \textbf{I}_{NL} + (\textbf{v}_k^H\textbf{v}_k)(\bm{\Sigma}_1 \otimes \textbf{I}_{NL}) \right) \tvec{\textbf{W}} \notag \\
&= \textbf{v}_k^H\textbf{v}_k + (\tvec{\textbf{W}})^H\left(\textbf{C}_k^T \otimes \textbf{I}_{NL}\right)\tvec{\textbf{W}}.
\end{align}
Define $\textbf{T}_k = (\Gcap^H\textbf{v}_k\textbf{v}_k^H\Gcap)^T \otimes (\hcap_k\hcap_k^H), \: \textbf{F}_{\ell,k}=\textbf{C}_{\ell}^T \otimes \textbf{D}_k, \: \textbf{E}_k=\textbf{I}_{NL} \otimes \textbf{D}_k, \: \textbf{J}_k = \textbf{C}_k^T \otimes \textbf{I}_{NL}$ for all $k$. Then we can write SINR conditions and total mean power as 
\begin{equation} \label{C3}
\begin{aligned}
(\tvec{\textbf{W}})^H\left[(1+\gamma_k)\textbf{T}_k-\gamma_k\displaystyle\sum_{\ell=1}^K\textbf{F}_{\ell,k}-\gamma_k\sigma_{\text{RRH}}^2\textbf{E}_k\right]\tvec{\textbf{W}} \geq \gamma_k\sigma_{\text{MS}}^2, \: \: \forall k \\
P = \displaystyle\sum_{k=1}^K \textbf{v}_k^H\textbf{v}_k + (\tvec{\textbf{W}})^H\left(\sigma_{\text{RRH}}^2\textbf{I}_{N^2L^2}+\displaystyle\sum_{k=1}^K\textbf{J}_k\right)\tvec{\textbf{W}}.
\end{aligned}
\end{equation}
The matrix $\textbf{W}$ is block diagonal and it includes $NL^2$ many unknowns. Other $(N^2-N)L^2$ entries are zero. There exists a matrix $\textbf{U}: \: N^2L^2 \times NL^2$ and a vector of unknown variables $\textbf{w}_0: \: NL^2 \times 1$ such that $\tvec{\textbf{W}} = \textbf{U} \textbf{w}_0$. Here, each column of $\textbf{U}$ includes a single 1 and other entries are equal to 0. We put the 1's at the entries corresponding to the unknown variables in $\tvec{\textbf{W}}$. After this observation, we can write the problem in terms of $\textbf{w}_0$:
\begin{equation} \label{C4}
\begin{aligned}
\textbf{w}_0^H\textbf{U}^H\left[(1+\gamma_k)\textbf{T}_k-\gamma_k\displaystyle\sum_{\ell=1}^K\textbf{F}_{\ell,k}-\gamma_k\sigma_{\text{RRH}}^2\textbf{E}_k\right]\textbf{U}\textbf{w}_0\geq \gamma_k\sigma_{\text{MS}}^2, \: \: \forall k \\
P = \displaystyle\sum_{k=1}^K \textbf{v}_k^H\textbf{v}_k + \textbf{w}_0^H\textbf{U}^H\left(\sigma_{\text{RRH}}^2\textbf{I}_{N^2L^2}+\displaystyle\sum_{k=1}^K\textbf{J}_k\right)\textbf{U}\textbf{w}_0.
\end{aligned}
\end{equation}
Finally we define $\bm{\mathcal{W}} = \textbf{w}_0\textbf{w}_0^H$ satisfying $\bm{\mathcal{W}} \succeq 0$ and $\text{rank}(\bm{\mathcal{W}})=1$. Using the variable $\bm{\mathcal{W}}$, we can formulate the problem as
\begin{equation} \label{C5}
\begin{aligned}
&(\text{P}1) \: \: \underset{\bm{\mathcal{W}}}{\min} \: \displaystyle\sum_{k=1}^K \textbf{v}_k^H\textbf{v}_k + \tr\left[\left(\textbf{U}^H\left(\sigma_{\text{RRH}}^2\textbf{I}_{N^2L^2}+\displaystyle\sum_{k=1}^K\textbf{J}_k\right)\textbf{U}\right)\bm{\mathcal{W}}\right] \\
&\text{such that} \: \tr\left[\left(\textbf{U}^H\left((1+\gamma_k)\textbf{T}_k-\gamma_k\displaystyle\sum_{\ell=1}^K\textbf{F}_{\ell,k}-\gamma_k\sigma_{\text{RRH}}^2\textbf{E}_k\right)\textbf{U}\right)\bm{\mathcal{W}} \right] \geq \gamma_k\sigma_{\text{MS}}^2, \: \: \forall k \\
&\bm{\mathcal{W}} \succeq 0, \: \text{rank}(\bm{\mathcal{W}})=1.
\end{aligned}
\end{equation}
In (P1), cost and all constraints except the rank constraint are convex. By omitting the rank constraint it can be solved with SDR using standard convex optimization tools such as SeDuMi \cite{SeDuMi}, CVX \cite{CVX}, Mosek \cite{Mosek}. 

\vspace{-5mm}

\subsection{Fronthaul Link Beamformer Design}
In this part, we consider the case where $\textbf{W}$ is fixed. In this case, we can write
\begin{equation} \label{C6}
\begin{aligned}
\left|\widehat{\textbf{h}}_k^H\textbf{W}\widehat{\textbf{G}}^H\textbf{v}_k\right|^2 &= \textbf{v}_k^H\Gcap\textbf{W}^H\hcap_k\hcap_k^H\textbf{W}\Gcap^H\textbf{v}_k \\
\tr\left(\textbf{D}_k\textbf{W}\textbf{C}_{\ell}\textbf{W}^H\right) &= \tr\left(\textbf{W}^H\textbf{D}_k\textbf{W}\left(\Gcap^H\textbf{v}_{\ell}\textbf{v}_{\ell}^H\Gcap+(\textbf{v}_{\ell}^H\textbf{v}_{\ell})\bm{\Sigma}_1\right)\right) \\
&= \textbf{v}_{\ell}^H\left[\Gcap\textbf{W}^H\textbf{D}_k\textbf{W}\Gcap^H+\tr(\textbf{W}^H\textbf{D}_k\textbf{W}\bm{\Sigma}_1)\textbf{I}_M\right]\textbf{v}_{\ell}. 
\end{aligned}
\end{equation}
Let $\textbf{A}_k=\Gcap\textbf{W}^H\hcap_k\hcap_k^H\textbf{W}\Gcap^H, \: \textbf{B}_k = \Gcap\textbf{W}^H\textbf{D}_k\textbf{W}\Gcap^H+\tr(\textbf{W}^H\textbf{D}_k\textbf{W}\bm{\Sigma}_1)\textbf{I}_M, \: \textbf{V}_k=\textbf{v}_k\textbf{v}_k^H, \: \forall k$ and $a=\sigma_{\text{RRH}}^2 \tr\left(\textbf{W}^H \textbf{W}\right) , \: b=\sigma_{\text{RRH}}^2\tr\left(\textbf{D}_k\textbf{W}\textbf{W}^H\right)+\sigma_{\text{MS}}^2$. Using (\ref{P_eqn}) and (\ref{C6}), we formulate the problem as 
\begin{equation} \label{C7}
\begin{aligned}
&(\text{P}2) \: \: \underset{\{\textbf{V}_k\}_{k=1}^K}{\min} \: \displaystyle\sum_{k=1}^K \tr\left(\bm{\tau_0} \textbf{V}_k\right) + a \\
&\text{such that} \: \dfrac{\tr\left(\textbf{A}_k\textbf{V}_k\right)}{\displaystyle\sum_{\ell=1}^K\tr\left(\textbf{B}_k\textbf{V}_{\ell}\right)-\tr\left(\textbf{A}_k\textbf{V}_k\right)+b} \geq \gamma_k, \: \: \forall k, \quad \textbf{V}_k \succeq 0, \: \text{rank}(\textbf{V}_k)=1, \: \: \forall k.
\end{aligned}
\end{equation}
(P2) can also be solved using convex optimization tools by omitting the rank constraints.

\vspace{-5mm}

\subsection{Rank-1 Approximation for SDR}
In both fronthaul and access link beamformer designs, we find a solution by omitting the rank constraint. If the result is rank-1, the solution becomes optimal. Otherwise, we apply a widely used randomization method \cite{AF-Relay1}-\cite{DF-2}, \cite{Ch-Add-1}. Let $\textbf{X}$ be the matrix found after convex optimization. We want to find a vector $\textbf{x}$ satisfying $\textbf{X}=\textbf{x}\textbf{x}^H$ which is not possible if $\text{rank}(\textbf{X})>1$. In such a case, we select $\textbf{x}=\textbf{E}\bm{\Lambda}^{1/2}\textbf{y}$ where $\textbf{X}=\textbf{E}\bm{\Lambda}\textbf{E}^H$ is the eigenvalue decomposition of $\textbf{X}$ and $\textbf{y}$ is a zero-mean real Gaussian random vector with unity covariance matrix.

\vspace{-5mm}
 
\subsection{Alternating Optimization (AO) Method}
We know that each one of fronthaul and access link beamformers can be found using convex optimization with SDR approach by fixing the other. Using this idea we can find a solution for (P0) by alternating optimization of fronthaul and access link beamformers. In general alternating optimization methods converge to local optimum points. The choice of initial point affects the performance. We consider the CP-to-RRH transmissions and use the total SNR at RRHs to find a suitable initial point. Let $\text{SNR}_{kn}$ be the SNR of $k$-th user at $n$-th RRH, i.e., $\text{SNR}_{kn} = \dfrac{\lVert \Gcap_n^H\textbf{v}_k \rVert^2}{\sigma_{\text{RRH}}^2}.$ The total SNR is given by $\text{SNR}_{\text{tot}} = \displaystyle\sum_{n=1}^N\displaystyle\sum_{k=1}^K \text{SNR}_{kn} = \tr\left(\textbf{V}^H\textbf{G}_0\textbf{V}\right)$ where $\textbf{G}_0 = \dfrac{1}{\sigma_{\text{RRH}}^2}\displaystyle\sum_{n=1}^N \Gcap_n\Gcap_n^H$ and $\textbf{V} = \left[ \textbf{v}_1 \: \textbf{v}_2 \: \cdots \: \textbf{v}_K \right]$. We know that $P_{\text{CP}}=\tr\left(\textbf{V}^H\textbf{V}\right)$. Furthermore, in order to send the user data from CP to RRHs properly, we need $M \geq K$ and $\text{rank}(\textbf{V})=K$. To satisfy these constraints, we choose $\textbf{V}$ such that $\textbf{V}^H\textbf{V}=\sqrt{\dfrac{P_{\text{CP}}}{K}} \textbf{I}_K$. We aim to find $\textbf{V}$ maximizing $\text{SNR}_{\text{tot}}$. By Von-Neumann's Inequality, we have
\begin{equation}
\tr\left(\textbf{V}\textbf{V}^H\textbf{G}_0 \right) \leq \displaystyle\sum_{i=1}^M \lambda_i\left(\textbf{V}\textbf{V}^H\right)\lambda_i\left(\textbf{G}_0 \right)
= \displaystyle\sum_{i=1}^K \lambda_i\left(\textbf{V}\textbf{V}^H\right)\lambda_i\left(\textbf{G}_0 \right)
= \dfrac{P_{\text{CP}}}{K}\displaystyle\sum_{i=1}^K \lambda_i\left(\textbf{G}_0 \right).
\end{equation}
Notice that the $K$ largest eigenvalues of $\textbf{V}\textbf{V}^H$ are equal to $\dfrac{P_{\text{CP}}}{K}$ and other $M-K$ are equal to zero. The equality holds when we have
\begin{equation} \label{Init} 
\textbf{v}_k = \sqrt{\dfrac{P_{\text{CP}}}{K}}e_k\left(\textbf{G}_0\right), \: \forall k.
\end{equation}
To find a suitable initial point we select the CP beamformers as in (\ref{Init}). On the other hand, the selection of initial $P_{\text{CP}}$ is also required. To perform this task, we use Algorithm 0. 

\vspace{-3mm}

\noindent\rule{\textwidth}{0.8pt} 

\vspace{-3mm}

\noindent\textbf{Algorithm 0} (Initialization for Alternating Optimization) \vspace{-5mm} \\
\noindent\rule{\textwidth}{0.4pt} \\
Set $P_{\text{CP}}^{(0)}=1, \: \mu_0 = 1.05, \: t_{\text{max}, 0}=100$. For $t=0, 1, 2, \ldots, t_{\text{max}, 0}$ repeat the following steps:
\begin{itemize}
\item Form $\textbf{v}_k^{(t)} = \sqrt{\dfrac{P_{\text{CP}}^{(t)}}{K}}e_k\left(\textbf{G}_0\right), \: \forall k$. Solve (P1) to find $\textbf{W}^{(t)}$. 
\item If the problem is feasible, then set the initial value of $\textbf{W}$ as $\textbf{W}^{(t)}$ and terminate. 
\item Set $P_{\text{CP}}^{(t+1)}=\mu_0 P_{\text{CP}}^{(t)}$. 
\end{itemize}
\vspace{-5mm}
\noindent\rule{\textwidth}{0.4pt}
Algorithm 0 is used to find the initial value of $\textbf{W}$. Starting from this value, we apply alternating optimization by solving (P1) and (P2) iteratively. At each iteration, $P$ decreases since both (P1) and (P2) minimizes $P$ when one of fronthaul and access link beamformers is fixed. As the power is limited below ($P \geq 0$) we conclude by Monotone Convergence Theorem \cite{MCT} that this method is convergent. When the rate of change of $P$ is small enough we stop the iteration and find the final solution. The method is summarized in Algorithm 1. 

\vspace{-3mm}

\noindent\rule{\textwidth}{0.8pt}

\vspace{-3mm}
 
\noindent\textbf{Algorithm 1} (Alternating Optimization) \vspace{-5mm} \\
\noindent\rule{\textwidth}{0.4pt} \\
Using Algorithm 0, find the initial value $\textbf{W}^{(0)}$. Define $t_{\text{max}, 1}=100, \: \eta=10^{-3}$. For $t=0, 1, \ldots, t_{\text{max}, 1}$, repeat the following steps:
\begin{itemize}
\item Solve (P2) to find $\textbf{v}_k^{(t)}, \: \forall k$. Solve (P1) to find $\textbf{W}^{(t)}$. 
\item If $|P^{(t)}-P^{(t-1)}|<\eta P^{(t)}$, then terminate. 
\end{itemize}
\vspace{-5mm}
\noindent\rule{\textwidth}{0.4pt} 

\vspace{-5mm}

\subsection{Total SNR Max (TSM) Method }
Algorithm 0 is used to find an initial point for AO method. By extending Algorithm 0, we propose another iterative method, called the Total SNR Max (TSM) Method, which is computationally less complex compared to AO. Firstly, we make an observation for values of $P$ as $P_{\text{CP}}$ increases. Assume that we use (\ref{Init}) to form CP beamformers. Starting from a small value, we increase $P_{\text{CP}}$ continuously and at each time we find the corresponding RRH beamforming matrix by solving (P1) as in Algorithm 0. We observe that in general there exist two iteration indices $0<t_1<t_2$ such that the problem is infeasible for $t<t_1$, $P^{(t)}$ is decreasing for $t_1<t<t_2$, and increasing for $t>t_2$. This shows that optimal value of $P$ is achieved when $t=t_2$. By the motivation of this observation, we propose Algorithm 2. 

\vspace{-3mm}

\noindent\rule{\textwidth}{0.8pt} 

\vspace{-3mm}

\noindent\textbf{Algorithm 2} (Total SNR Max Method) \vspace{-5mm} \\
\noindent\rule{\textwidth}{0.4pt} \\
Set $P_{\text{CP}}^{(0)}=1, \: P^{(0)}=0, \: \mu_2=1.05$. For $t=0, 1, \ldots, t_{\text{max}, 2}=100$, repeat the following steps:
\begin{itemize}
\item Form $\textbf{v}_k^{(t)}=\sqrt{\dfrac{P_{\text{CP}}^{(t)}}{K}}e_k\left(\textbf{G}_0\right), \: \forall k$.
\item Solve (P1) to find $\textbf{W}^{(t)}$. If the problem is feasible then evaluate $P^{(t)}$ using CP and RRH beamformers. Otherwise, set $P^{(t)}=0$. 
\item If $P^{(t)}>P^{(t-1)}>0$, then terminate. 
\item Set $P_{\text{CP}}^{(t+1)}=\mu_2 P_{\text{CP}}^{(t)}$. 
\end{itemize}
\vspace{-5mm}
\noindent\rule{\textwidth}{0.4pt}
This algorithm finds CP beamformers using the approach given in (\ref{Init}) by iteratively changing the $P_{\text{CP}}$ value. RRH beamforming selection is done as in AO method. 

\vspace{-5mm}

\subsection{Complexity of Convex Optimization Methods}
In general, we can measure the computational complexity of AO and TSM as the product of number of iterations and the complexity at each iteration. At each iteration, the main component of complexity is related to the convex optimization and all other operations can be neglected. We use SeduMi as the convex optimization tool to implement AO and TSM. In both methods, at each iteration, we minimize $c^Hx$ subject to $Ax=b$ where $x \in \mathbb{C}^{n}$ is the vector of all unknowns and $A \in \mathbb{C}^{m \times n}, \: b\in\mathbb{C}^m, \: c\in \mathbb{C}^{n}$ are known vectors/matrices. We know by \cite{SeDuMi} that the corresponding computational complexity is $\mathcal{O}(n^2m^{2.5}+m^{3.5})$ for SeDuMi. The corresponding $m$ and $n$ values for fronthaul and access link beamforming designs are calculated as 
\begin{equation}
\text{Fronthaul Link}: \: \: m=K, \: \: n=K+KM^2, \: \: \text{Access Link}: \: \: m=K, \: \: n=K+N^2L^4. 
\end{equation}
In AO, both fronthaul and access link beamformer designs are done by convex optimization, meanwhile, TSM uses convex optimization only for access link. Hence, the corresponding computational complexities are given by
\begin{equation}
\begin{aligned}
\text{Complexity of AO}: &\quad \mathcal{O}\left(N_{\text{AO}}K^{2.5}\left[(K+KM^2)^2+(K+N^2L^4)^2+2K\right]\right), \\
\text{Complexity of TSM}: &\quad \mathcal{O}\left(N_{\text{TSM}}K^{2.5}\left[(K+N^2L^4)^2+K\right]\right)
\end{aligned}
\end{equation}
where $N_{\text{AO}}$ and $N_{\text{TSM}}$ are number of iterations for AO and TSM, respectively. In simulation results, we show that the number of iterations for both methods are similar and average complexity of AO is larger than that of TSM, as expected. 
\vspace{-4mm}
\section{Standard Beamforming Methods}
In this section, we present two algorithms adapted from well-known beamforming methods. These methods are based on MRC, ZF and SVD. The purpose of considering these methods is to observe the performance of well-known methods in our joint beamforming design problem. We also make a comparison with the performance bound and relatively complex convex optimization methods described in the previous section. In the first method, called MRC-ZF, we design fronthaul beamformers using the MRC idea. Access link beamformers are chosen as in ZF to cancel the interference due to other user signals. The second method is called SVD-ZF where the fronthaul beamformers are designed by an SVD algorithm. The access link beamformers are chosen to cancel the interference as in MRC-ZF. Because of the nature of the problem, a direct implementation is not possible. We need some adaptations to use MRC, ZF, and SVD. 
\vspace{-5mm}
\subsection{MRC-ZF}
We know that MRC optimizes the signal power by a coherent reception. ZF eliminates the interference and hence enhances the SINR. By the motivation of these beamforming methods, we choose the fronthaul and access link beamformers as
\begin{equation} \label{S1}
\textbf{v}_k = \left(\hcap_k^H\textbf{W}\Gcap^H\right)^H, \: \: \forall k, \quad \hcap_k^H\textbf{W}\Gcap^H\textbf{v}_{\ell} = \delta[k-\ell], \: \: \forall k, \ell. 
\end{equation}
In this method, $\textbf{v}_k$'s are chosen as the conjugate-transpose of the corresponding effective channel $\hcap_k^H\textbf{W}\Gcap^H$. The matrix $\textbf{W}$ is chosen to cancel the interference due to undesired user signals. Notice that both beamformers are chosen in terms of channel estimates only. This approach is used to make the algorithm simpler. Using (\ref{S1}), we get  
\begin{equation} 
\hcap_k^H\textbf{W}\Gcap^H\Gcap\textbf{W}^H\hcap_{\ell}=\tr\left(\textbf{W}^H\hcap_{\ell}\hcap_k^H\textbf{W}\Gcap^H\Gcap\right)=\delta[k-\ell] , \: \: \forall k, \ell. 
\end{equation}
Using the fact that $\tvec{\textbf{W}}=\textbf{U}\textbf{w}_0$, we obtain that 
\begin{equation} \label{S2}
\textbf{w}_0^H \textbf{U}^H\left[(\Gcap^H\Gcap)^T \otimes (\hcap_{\ell}\hcap_k^H)\right]\textbf{U}\textbf{w}_0=\delta[k-\ell], \: \: \forall k, \ell. 
\end{equation}
(\ref{S2}) is a quadratically constrained quadratic program (QCQP) type problem including a set of second order matrix equations with $NL^2$ unknowns and $K^2$ equations. If $NL^2 \geq K^2$, then we can find a solution using a standard QCQP solver. Let $\textbf{W}_0$ and $\{\textbf{v}_{k,0}\}_{k=1}^K$ be some solutions of (\ref{S1}). We use $\textbf{v}_k=\sqrt{a}\textbf{v}_{k,0}, \: \forall k$ and $\textbf{W}=\sqrt{b}\textbf{W}_0$ where $a$ and $b$ are two non-negative real numbers. We use $a$ and $b$ to optimize the power allocation and minimize the total power spent. Using the beamformer expressions, we can write SINR constraints and total mean power as 
\vspace{-5mm}
\begin{equation} \label{S3}
\dfrac{ab\cdot c_{k,1}}{ab\cdot c_{k,2}-ab\cdot c_{k,1}+b\cdot c_{k,3}+c_{k,4}} \geq \gamma_k, \: \: \forall k, \: \: P = a\cdot d_5+ab\cdot d_6+b\cdot d_7
\end{equation}
where
\begin{equation}
\begin{aligned}
c_{k,1}&=\left|\widehat{\textbf{h}}_k^H\textbf{W}_0\widehat{\textbf{G}}^H\textbf{v}_{k,0}\right|^2, \: \: c_{k,2}=\displaystyle\sum_{\ell=1}^K\tr\left(\textbf{D}_k\textbf{W}_0\left(\Gcap^H\textbf{v}_{\ell,0}\textbf{v}_{\ell,0}^H\Gcap+(\textbf{v}_{\ell,0}^H\textbf{v}_{\ell,0})\bm{\Sigma}_1\right)\textbf{W}_0^H\right) \\
c_{k,3}&=\sigma_{\text{RRH}}^2\tr\left(\textbf{D}_k\textbf{W}_0\textbf{W}_0^H\right), \: \: c_{k,4}=\sigma_{\text{RRH}}^2, \: \: d_5=\displaystyle\sum_{k=1}^K \textbf{v}_{k,0}^H\textbf{v}_{k,0} \\
d_6 &= \displaystyle\sum_{k=1}^K \textbf{v}_{k,0}^H\Gcap\textbf{W}_0^H\textbf{W}_0\Gcap^H\textbf{v}_{k,0} + \tr\left(\textbf{W}_0^H\textbf{W}_0\bm{\Sigma}_1\right)\displaystyle\sum_{k=1}^K \textbf{v}_{k,0}^H\textbf{v}_{k,0}, \: \: d_7 = \sigma_{\text{RRH}}^2\tr\left(\textbf{W}_0^H\textbf{W}_0\right).
\end{aligned}
\end{equation}
Using the SINR constaints in (\ref{S3}), we get 
\begin{equation} \label{S4}
a \geq d_{k,1}+\dfrac{d_{k,2}}{b}, \: \: (1+\gamma_k)c_{k,1}>\gamma_kc_{k,2}, \: \: \forall k
\end{equation}
where $d_{k,1}=\dfrac{\gamma_kc_{k,3}}{(1+\gamma_k)c_{k,1}-\gamma_kc_{k,2}}, \: d_{k,2}=\dfrac{\gamma_kc_{k,4}}{(1+\gamma_k)c_{k,1}-\gamma_kc_{k,2}}, \: \forall k$. The first condition in (\ref{S4}) provides $K$ inequalities for $a$ and $b$. The second condition should be satisfied to obtain a feasible solution. The problem of minimizing $P$ in (\ref{S3}) under SINR constraints given by (\ref{S4}) is a two-variable QCQP problem which can be solved directly. The solution steps are explained in Algorithm 3. 

\vspace{-3mm}

\noindent\rule{\textwidth}{0.8pt} 

\vspace{-3mm}

\noindent\textbf{Algorithm 3} (MRC-ZF) \vspace{-5mm} \\
\noindent\rule{\textwidth}{0.4pt}
\begin{itemize}
\item Find $\textbf{W}_0$ and $\{\textbf{v}_{k,0}\}_{k=1}^K$ by solving (\ref{S2}) using a QCQP solver. 
\item Check the feasibility condition given by (\ref{S4}). If it is not satisfied, then terminate.
\item For all $k$ evaluate $d_{k,1}, d_{k,2}, d_5, d_6, d_7$ using $\textbf{W}_0$ and $\{\textbf{v}_{k,0}\}_{k=1}^K$.
\end{itemize}
For each $k=1, 2, \ldots, K$ repeat the following steps:
\begin{itemize}
\item Find the solution interval $[b_1, b_2] \subseteq [0, \infty)$ of $b$ satisfying $d_{k,1}+\dfrac{d_{k,2}}{b} \geq d_{\ell,1}+\dfrac{d_{\ell,2}}{b}, \: \: \forall \ell \neq k$.
\item Evaluate the minimum value $P_{k,0}$ of $P = a\cdot d_5+ab\cdot d_6+b\cdot d_7$ for $a=d_{k,1}+\dfrac{d_{k,2}}{b}$ which is given by $P_{k,0}=d_{k,1}d_5+d_{k,2}d_6+d_7+2\sqrt{d_{k,1}d_{k,2}d_5d_6}$.
\item Evaluate the values of $P = a\cdot d_5+ab\cdot d_6+b\cdot d_7$ for $a=d_{k,1}+\dfrac{d_{k,2}}{b_1}, \: \: b=b_1$ and $a=d_{k,1}+\dfrac{d_{k,2}}{b_2}, \: \: b=b_2$ as $P_{k,1}$ and $P_{k,2}$. 
\item Evaluate the global minimum candidate for $k$ as $P_{\text{min},k}=\min (P_{k,0}, P_{k,1}, P_{k,2})$. 
\end{itemize}
Find the solution as $P_{\text{min}}=\min\limits_{k} P_{\text{min},k}$.

\vspace{-3mm}
\noindent\rule{\textwidth}{0.4pt}

Algorithm 3 optimally solves the beamforming design problem defined by MRC-ZF method. Notice that there is a feasibility condition defined by (\ref{S4}) which has to be satisfied in order to find a suitable beamformer. By the design method, the algorithm cancels the interference due to undesired user signals. As it uses the channel estimates only, the interference due to channel mismatch part cannot be canceled. The channel estimation error should be small enough to satisfy the feasibility condition. There is also another condition $NL^2 \geq K^2$ to find a solution for the matrix equation in (\ref{S2}). These conditions imply that MRC-ZF can be used if the channel estimation quality is good enough and the number of users is small enough. 
\vspace{-5mm}
\subsection{SVD-ZF} 
In TSM method, fronthaul beamformers are designed by maximizing the total SNR at RRHs. We have shown that the corresponding beamformer is found using SVD of a sum of channel components related to CP-to-RRH channels. We use this approach to design fronthaul beamformers and access link beamformers are found as in MRC-ZF. By the motivation of the SVD and ZF type operations, we name this method as SVD-ZF. We first consider the system of equations
\begin{equation} \label{S5}
\textbf{v}_k = e_k\left(\textbf{G}_0\right), \: \: \forall k, \quad \hcap_k^H\textbf{W}\Gcap^H\textbf{v}_{\ell} = \delta[k-\ell], \: \: \forall k, \ell
\end{equation}
where $\textbf{G}_0 = \dfrac{1}{\sigma_{\text{RRH}}^2}\displaystyle\sum_{n=1}^N \Gcap_n\Gcap_n^H$. The first condition in (\ref{S5}) maximizes the total SNR at RRHs and the second condition eliminates the interference. As in MRC-ZF, we only use channel estimates to design beamformers for simplicity. Using the transformation $\tvec{\textbf{W}}=\textbf{U}\textbf{w}_0$, we obtain that 
\begin{equation} \label{S6}
\hcap_k^H\textbf{W}\Gcap^He_{\ell}\left(\textbf{G}_0\right) = \left(\tvec{[\Gcap^He_{\ell}\left(\textbf{G}_0\right)\hcap_k^H]^T}\right)^T\textbf{U}\textbf{w}_0 \\
=\delta[k-\ell], \: \: \forall k, \ell.
\end{equation}
(\ref{S6}) includes a system of linear equations with $NL^2$ unknowns and $K^2$ equations. For $NL^2 \geq K^2$, we can find a solution using generalized matrix inversion.  As in MRC-ZF we optimize the power allocation to minimize the total power spent. Let $\textbf{W}_0$ and $\{\textbf{v}_{k,0}\}_{k=1}^K$ be some solutions of (\ref{S5}). We use $\textbf{v}_k=\sqrt{a}\textbf{v}_{k,0}, \: \forall k$ and $\textbf{W}=\sqrt{b}\textbf{W}_0$ where $a$ and $b$ are two non-negative real numbers. After this point, we can formulate the problem in terms of $a$ and $b$ as in MRC-ZF and find the optimal values following the same procedure. SVD-ZF is summarized in Algorithm 4. 

\vspace{-3mm}

\noindent\rule{\textwidth}{0.8pt} 

\vspace{-3mm}

\noindent\textbf{Algorithm 4} (SVD-ZF) \vspace{-5mm} \\
\noindent\rule{\textwidth}{0.4pt}
\begin{itemize}
\item Find $\textbf{W}_0$ and $\{\textbf{v}_{k,0}\}_{k=1}^K$ by solving (\ref{S6}) using generalized matrix inversion.
\item Apply the same procedure done in MRC-ZF to find the solution. 
\end{itemize}
\vspace{-5mm}
\noindent\rule{\textwidth}{0.4pt}
As in MRC-ZF, this method includes a feasibility condition including $\textbf{W}_0$ and $\{\textbf{v}_{k,0}\}_{k=1}^K$. We also need $NL^2 \geq K^2$ to find a solution for (\ref{S6}). Hence SVD-ZF also requires a good channel estimation quality and relatively small number of users. One can say that SVD-ZF is computationally less complex compared to MRC-ZF as it does not require a QCQP solver. 
\vspace{-4mm}
\section{Numerical Results}
In this section we compare the performances of the proposed methods with the performance bound by Monte Carlo simulations. Throughout the simulations, we assume that $\gamma_k = \gamma, \: \forall k$. We use a realistic channel model including path-loss, shadowing and small-scale fading defined in a 3GPP standard \cite{3GPP}. We consider a circular region in which CP is at the center, RRHs and MSs are distributed uniformly.\footnote{We choose the configurations where CP-to-RRH, CP-to-MS and RRH-to-MS distances are all at least 50 meters.} In Table I, the model parameters are presented. 

\vspace{-4mm}

{\renewcommand{\arraystretch}{0.7}
\begin{table}[ht]
\caption{Model parameters used in simulations}
\centering
\begin{tabular}{| c | c |}
\hline
Cell radius & $1$ km \\
\hline
Path-loss for Fronthaul Link ($P_{L,1}$) & $P_{L,1}=24.6+39.1\log_{10}d$ where $d$ is in meters \\
\hline
Path-loss for Access Link ($P_{L,2}$) & $P_{L,2}=36.8+36.7\log_{10}d$ where $d$ is in meters \\
\hline
Antenna gain (CP, RRH, MS) & $(9, 0, 0)$ dBi \\
\hline
Noise Figure (RRH, MS) & $(2, 10)$ dB \\
\hline
Bandwidth & 10 MHz \\
\hline
Noise power spectral density & $-174$ dBm/Hz \\
\hline
Small-scale fading model & Rayleigh, $\mathcal{C}\mathcal{N}(\textbf{0},\textbf{I})$ \\
\hline
Log-normal shadowing variance (CP, RRH) & $(6, 4)$ dB \\
\hline
\end{tabular}
\end{table}
}
\vspace{-4mm}
To generate channel estimates and channel estimation errors, we assume that pilot signal powers are adjusted according to the channel amplitudes so that the power ratios of $\mathbb{E}(|\Delta\textbf{G}_n|^2)/|\textbf{G}_n|^2$ and $\mathbb{E}(|\Delta\textbf{h}_{kn}|^2)/|\textbf{h}_{kn}|^2, \: \forall n, k$ are all equal to some known constant $\gamma_{\text{ch}}$. Here $\gamma_{\text{ch}}$ is a measure of channel estimation quality. Using the channel estimates and $\gamma_{\text{ch}}$, one can evaluate $\sigma_{1,n}^2, \: \forall n$ and $\sigma_{2,k,n}^2, \: \forall n, k$ accordingly. In simulations, we observe the effect of parameters $\gamma, K, N, L, M, \gamma_{\text{ch}}$. We know that there is always a non-zero probability of having an infeasible solution. To measure the ratio of feasibility, we define $P_{\text{success}}$ showing the percentage of feasible designs. We run $100$ Monte Carlo trials in each case. To evaluate the $P$ values for a method, we average the results over Monte Carlo trials with feasible solutions. 
\vspace{-4mm}
\begin{figure}[H]
\centering
\captionsetup{justification=centering}
\includegraphics[width=0.32\textwidth]{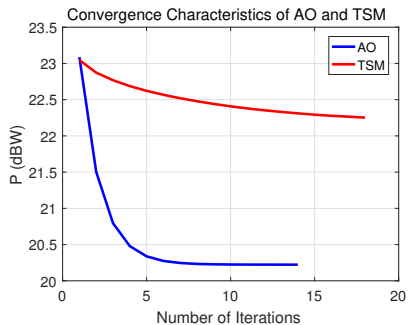}
\caption{Convergence characteristics of AO and TSM.}
\end{figure}
\vspace{-6mm}
In Fig. 2, we see a typical convergence graph of AO and TSM for $(K, N, L, M)=(4, 4, 4, 8)$, $\gamma=5$ dB and $\gamma_{\text{ch}}=0.01$. We observe that $P$ values decrease smoothly and both algorithms obtain a solution after a few iterations. 
\vspace{-4mm}   
\begin{figure}[H]
\centering
\captionsetup{justification=centering}
\subfloat{{\includegraphics[width=0.27\textwidth]{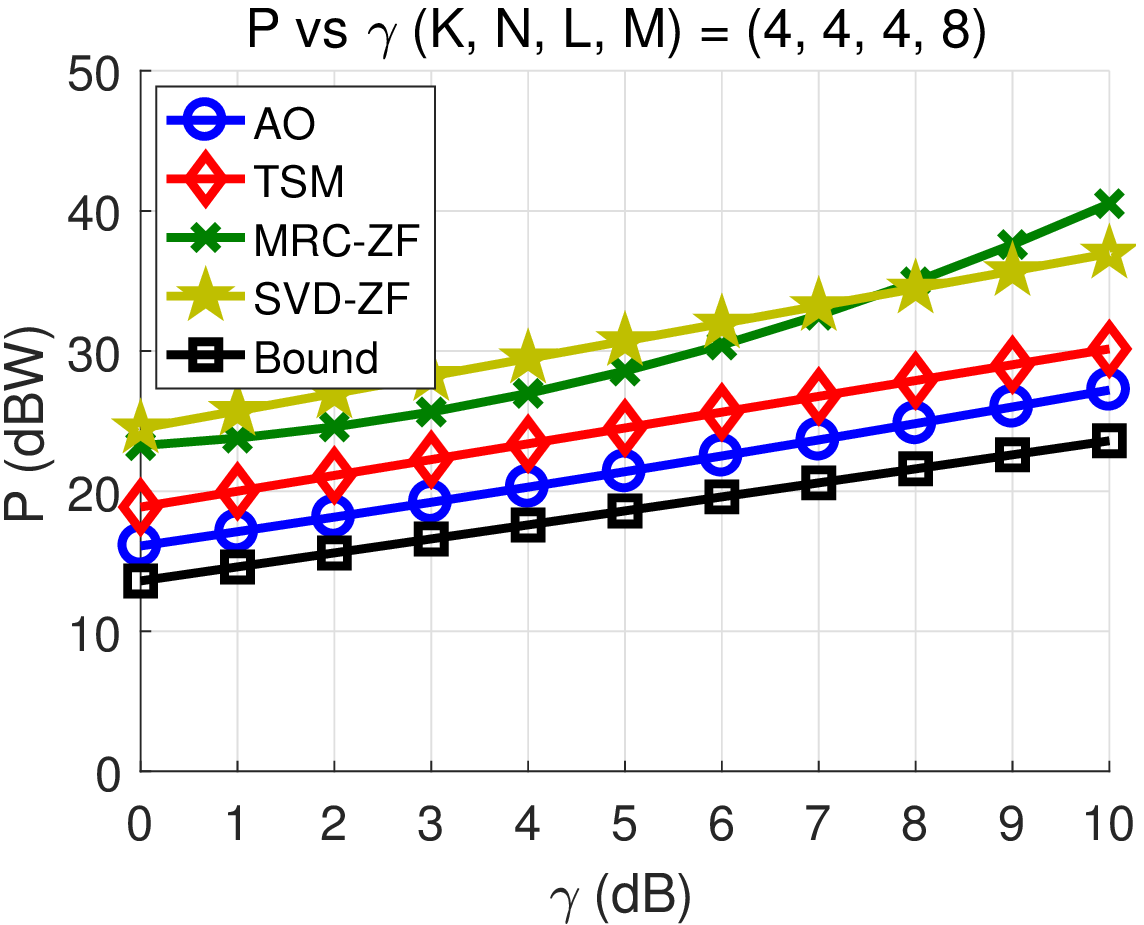}}}
\qquad
\subfloat{{\includegraphics[width=0.27\textwidth]{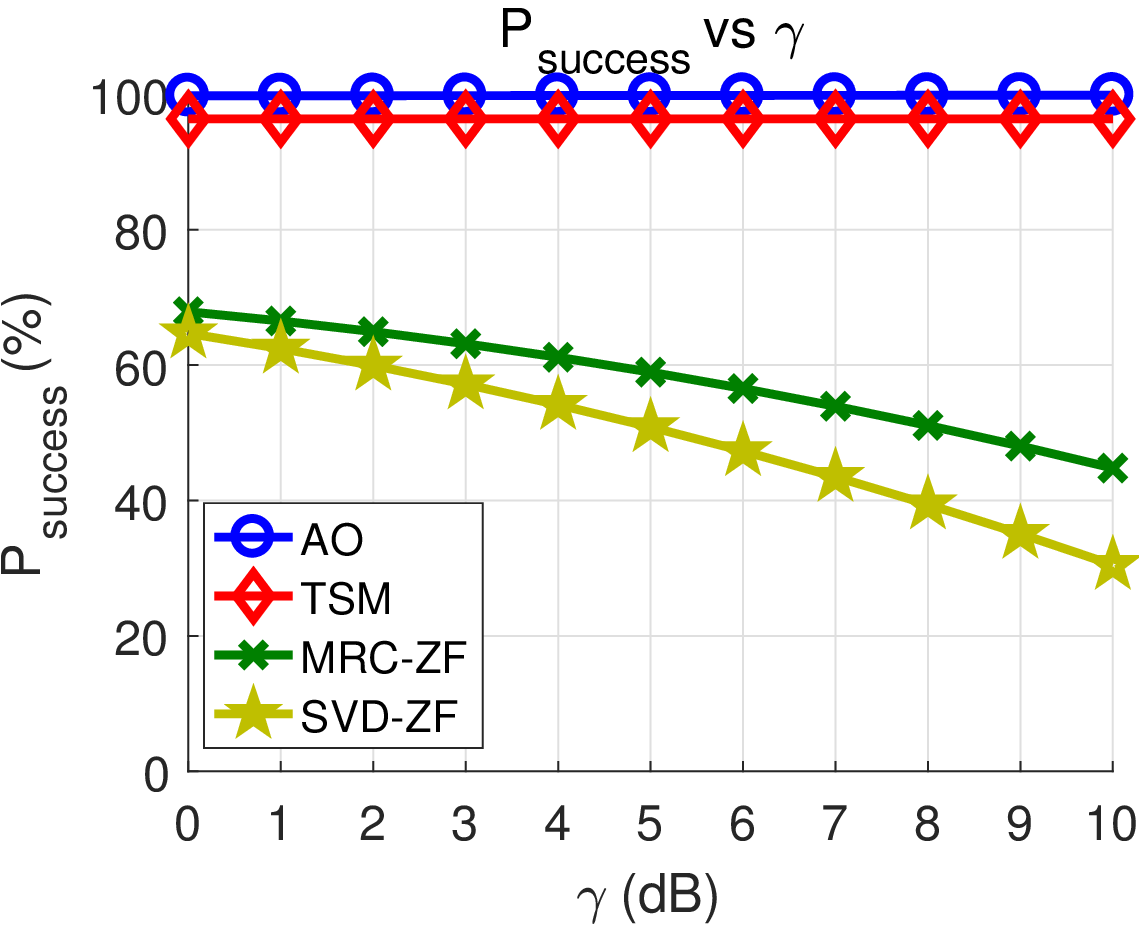}}}
\caption{$P$ and $P_{\text{success}}$ vs $\gamma$. $(K, N, L, M)=(4, 4, 4, 8), \: \gamma_{\text{ch}}=0.01$.}
\end{figure}
\vspace{-7mm}
In Fig. 3, we observe the effect of SINR threshold $\gamma$. For all $\gamma$ values, the performance loss compared to the bound are roughly $3$ and $6$ dB for AO and TSM, respectively. MRC-ZF and SVD-ZF have significantly worse performance than those of convex optimization methods. We observe that $P_{\text{success}}$ values of both methods decrease with $\gamma$. Even when $\gamma=0$ dB, infeasibility ratio is about $30$ percent for both methods. The results imply that even for a relatively low channel estimation error, the methods MRC-ZF and SVD-ZF may fail to solve the joint beamforming design problem with a large probability. We observe that AO can solve the problem with almost 100 percent whereas TSM feasibility ratio is slightly smaller than that of AO. 
\vspace{-4mm}
\begin{figure}[H]
\centering
\captionsetup{justification=centering}
\subfloat{{\includegraphics[width=0.27\textwidth]{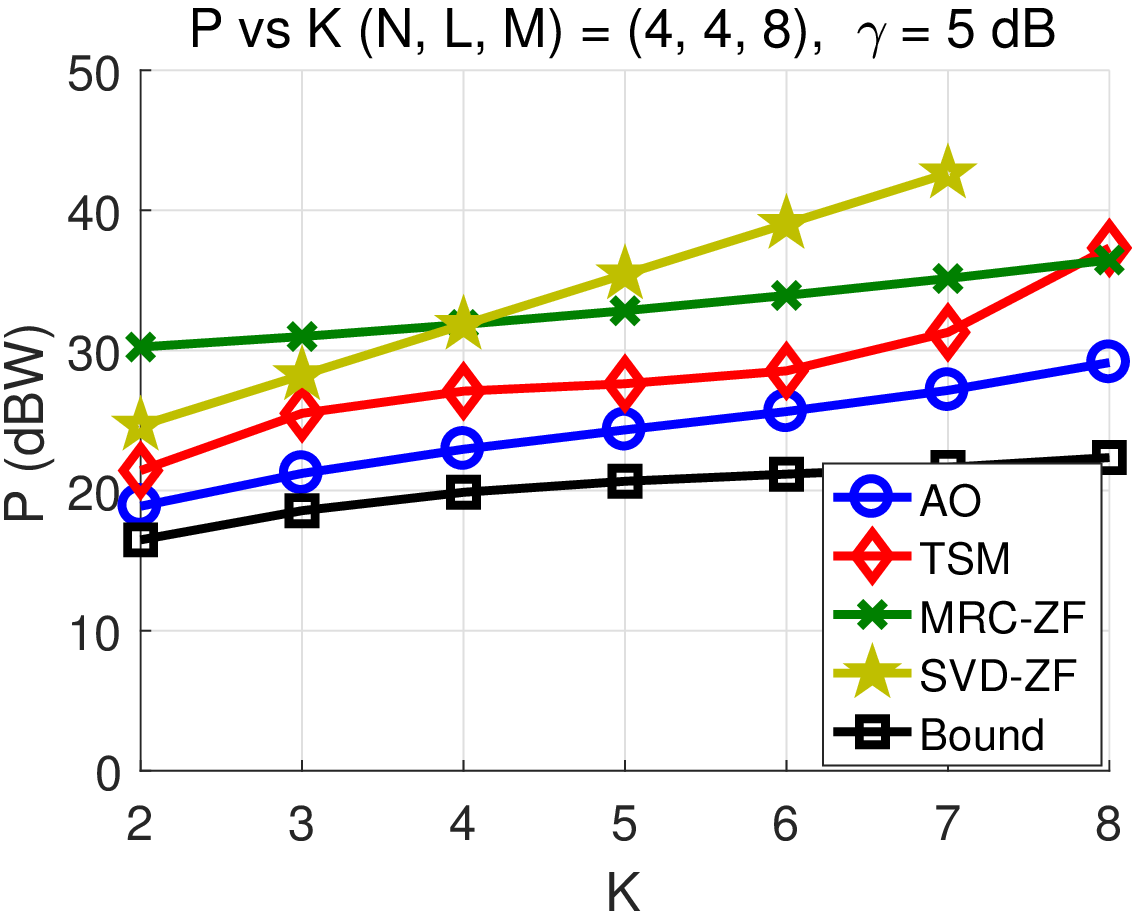}}}
\qquad
\subfloat{{\includegraphics[width=0.27\textwidth]{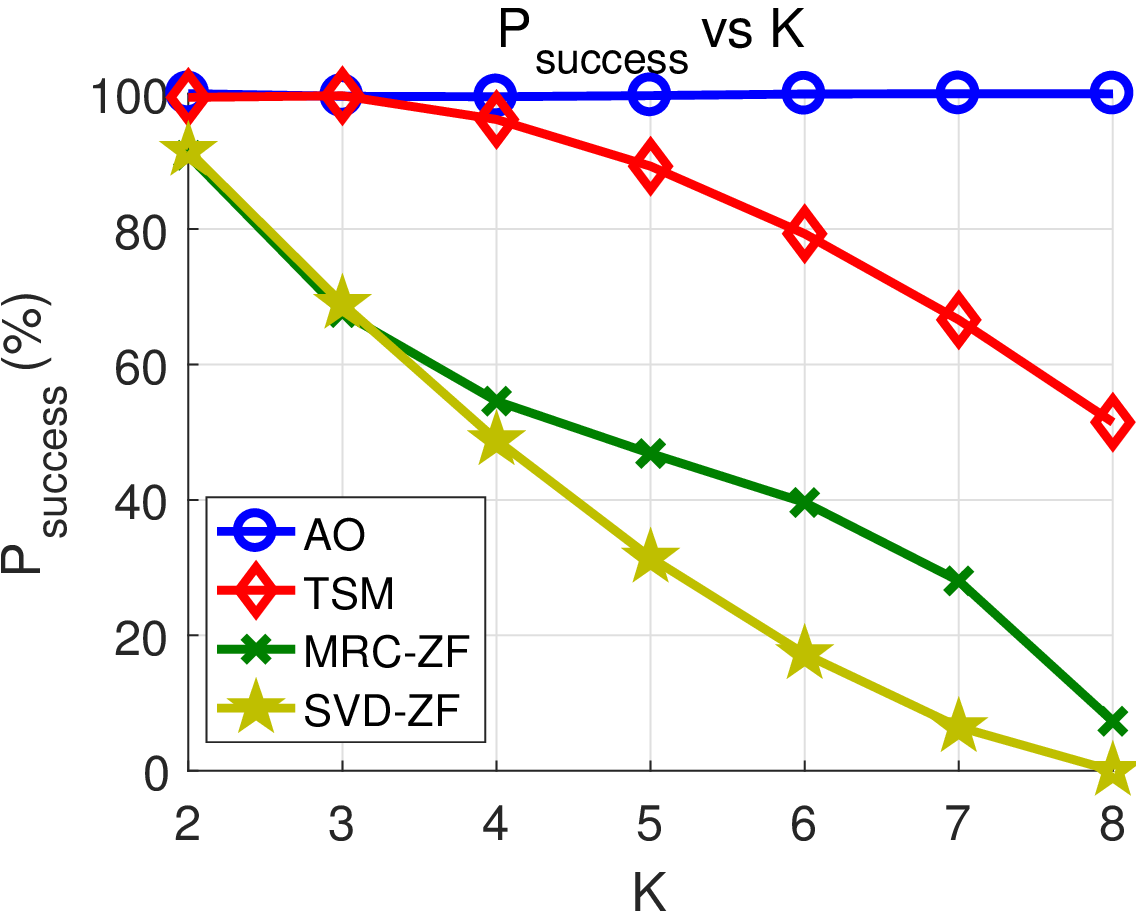}}}
\caption{$P$ and $P_{\text{success}}$ vs $K$. $(N, L, M)=(4, 4, 8), \: \gamma=5$ dB, $\gamma_{\text{ch}}=0.01$.}
\end{figure}
\vspace{-7mm}
Fig. 4 shows the performances as the number of MSs $K$ varies. We observe that the performance loss of all methods compared to the bound increase with $K$. This is due to the fact that bound can only be achieved when the interference due to undesired users is completely eliminated which becomes harder as $K$ increases. We see that for large $K$ values, the feasibility ratios of MRC-ZF and SVD-ZF become very small meaning that these methods cannot be used when the number of users is not small enough. Although it outperforms MRC-ZF and SVD-ZF, TSM performance also degrades for large number of users. On the other hand, AO can successfully design beamformers with 100 percent feasibility and it requires less power for all $K$ values compared to other three methods. 
\vspace{-4mm}
\begin{figure}[H]
\centering
\captionsetup{justification=centering}
\subfloat{{\includegraphics[width=0.30\textwidth]{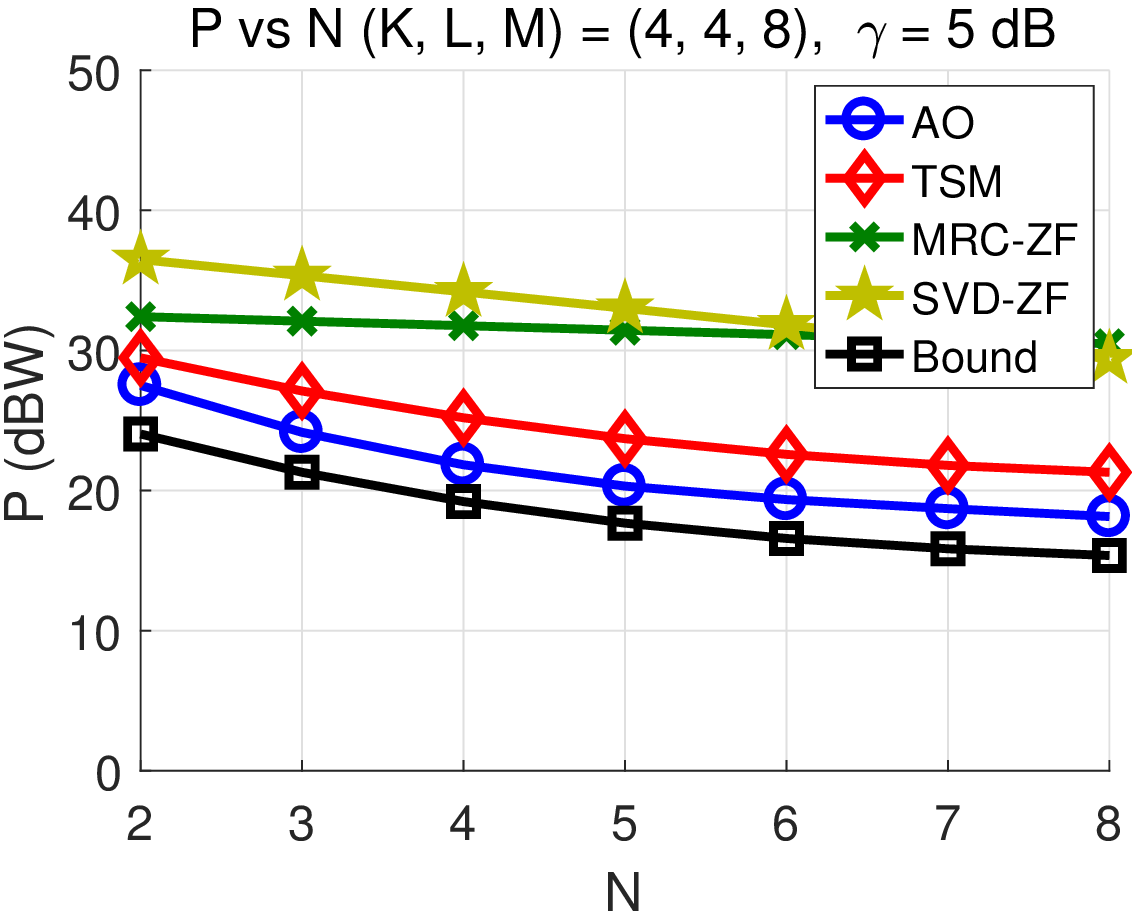}}}
\qquad
\subfloat{{\includegraphics[width=0.30\textwidth]{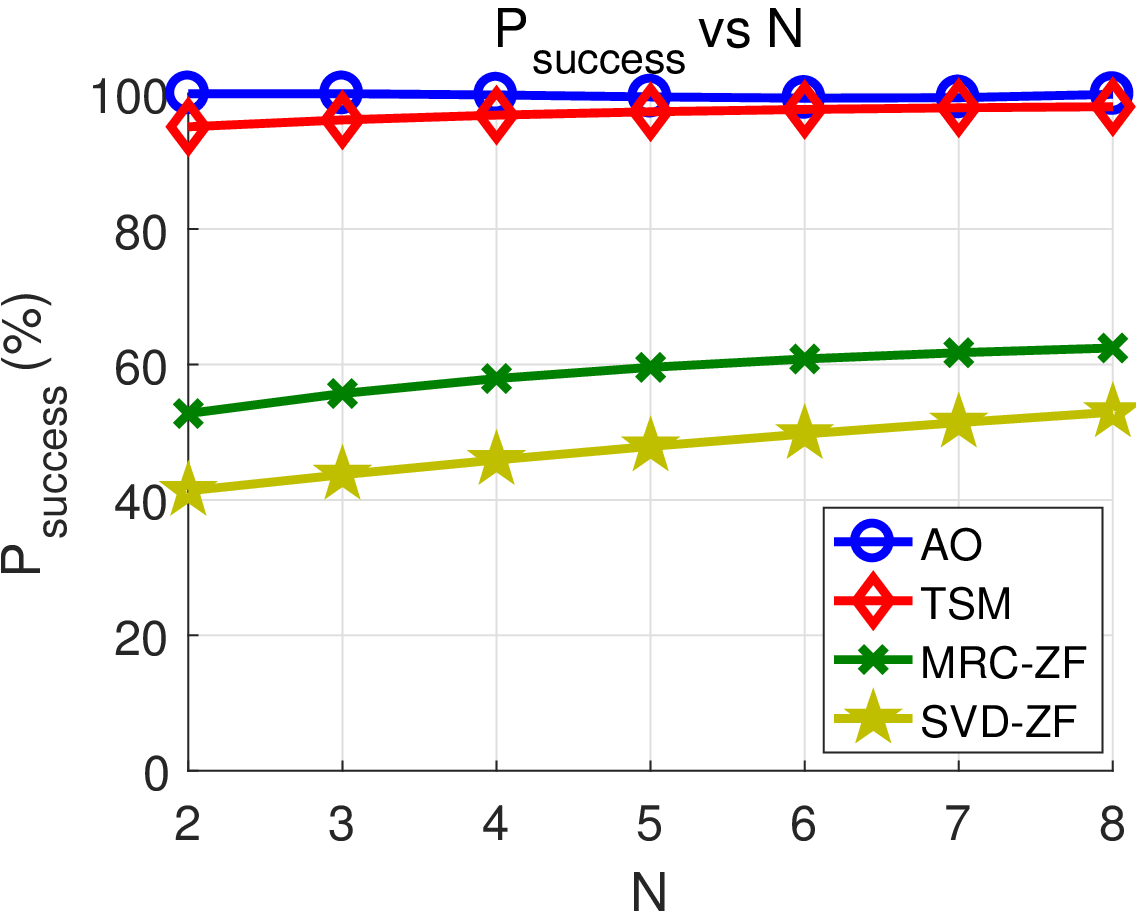}}}
\caption{$P$ and $P_{\text{success}}$ vs $N$. $(K, L, M)=(4, 4, 8), \: \gamma=5$ dB, $\gamma_{\text{ch}}=0.01$.}
\end{figure}
\vspace{-7mm}
\begin{figure}[H]
\centering
\captionsetup{justification=centering}
\subfloat{{\includegraphics[width=0.30\textwidth]{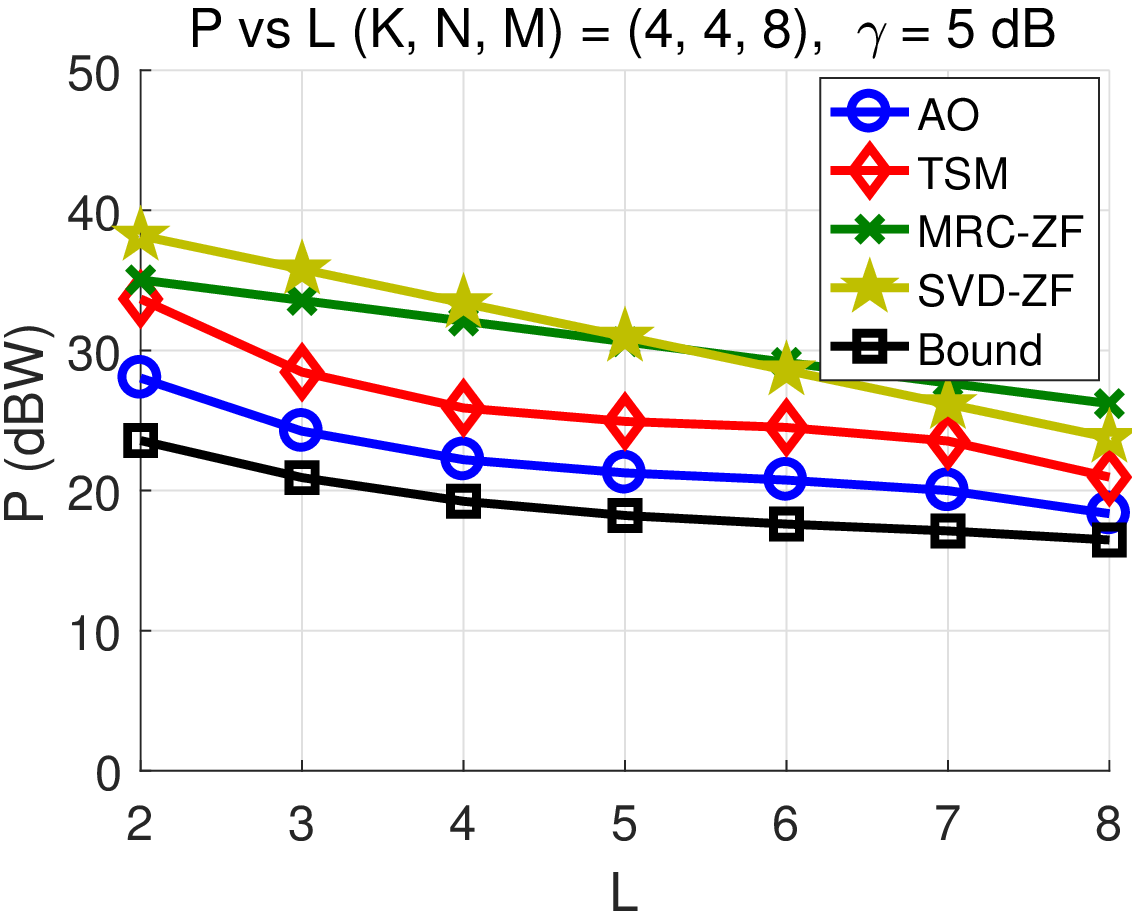}}}
\qquad
\subfloat{{\includegraphics[width=0.30\textwidth]{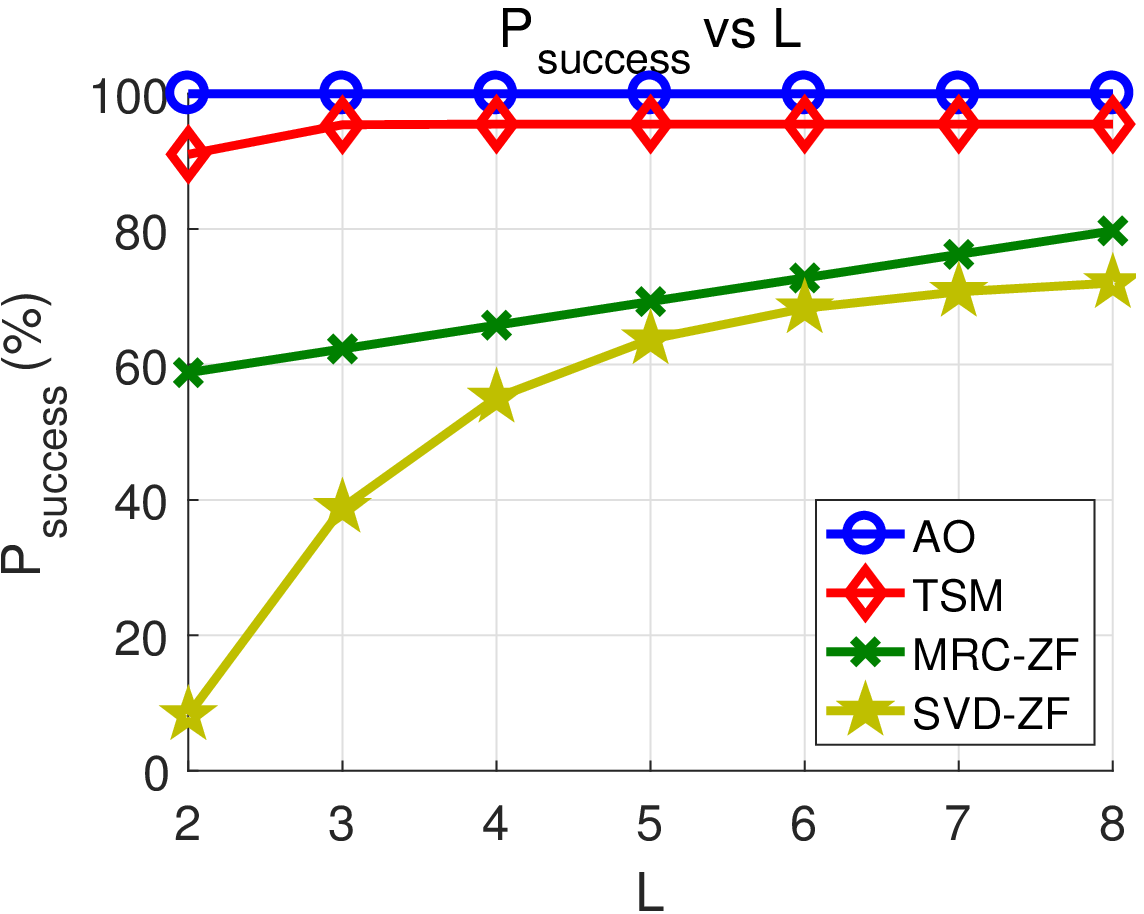}}}
\caption{$P$ and $P_{\text{success}}$ vs $L$. $(K, N, M)=(4, 4, 8), \: \gamma=5$ dB, $\gamma_{\text{ch}}=0.01$.}
\end{figure}   
\vspace{-7mm}
In Fig. 5-6, we observe the effects of the number of RRHs $N$ and the number of RRH antennas $L$. The results show that AO has the best performance for all cases. Its feasibility ratio is always 100 percent in these two simulations and the power difference with the bound is generally less than $5$ dB. The difference becomes smaller as $N$ or $L$ increases. As in the previous cases, MRC-ZF performs better than SVD-ZF and worse than TSM. We also observe that there is a significant difference in the bound values between $N=2$ and $N=8$ and the same fact is true for $L=2$ and $L=8$.  
\vspace{-4mm}
\begin{figure}[H]
\centering
\captionsetup{justification=centering}
\subfloat{{\includegraphics[width=0.30\textwidth]{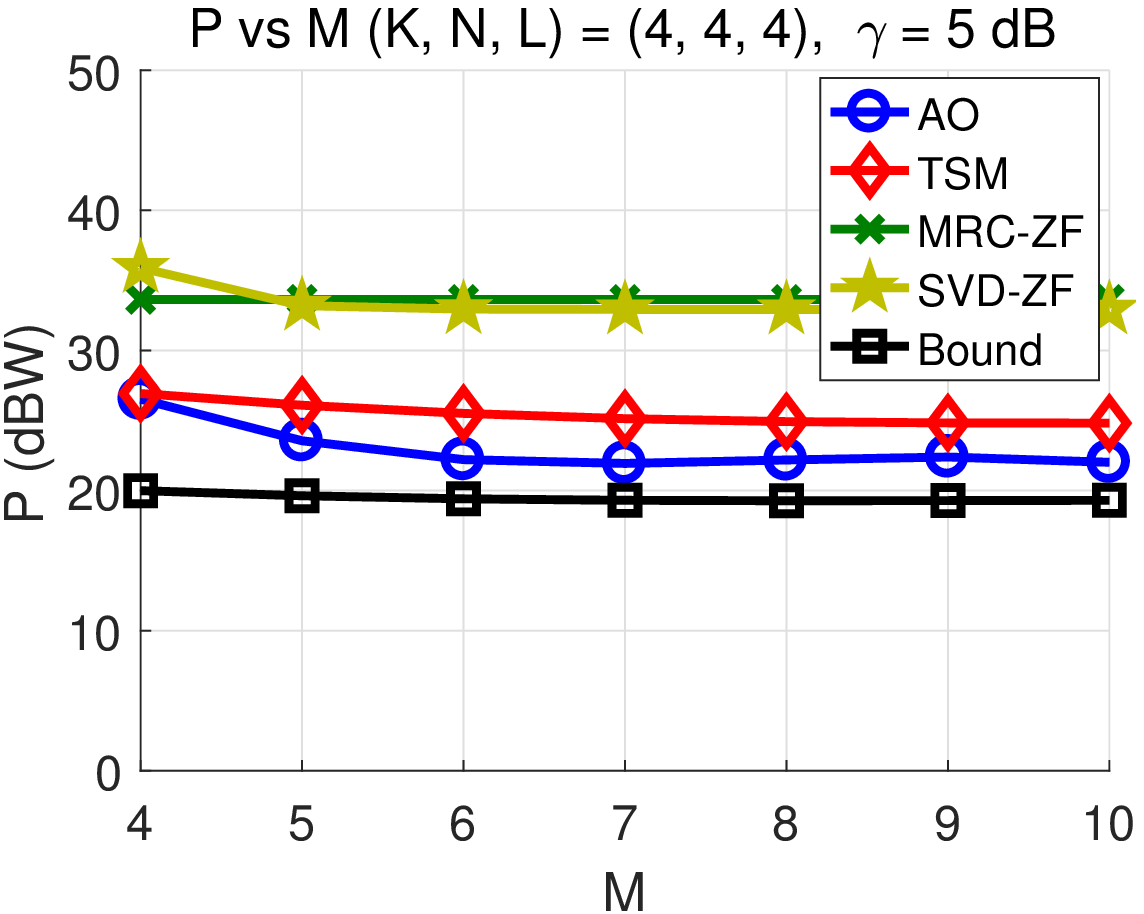}}}
\qquad
\subfloat{{\includegraphics[width=0.30\textwidth]{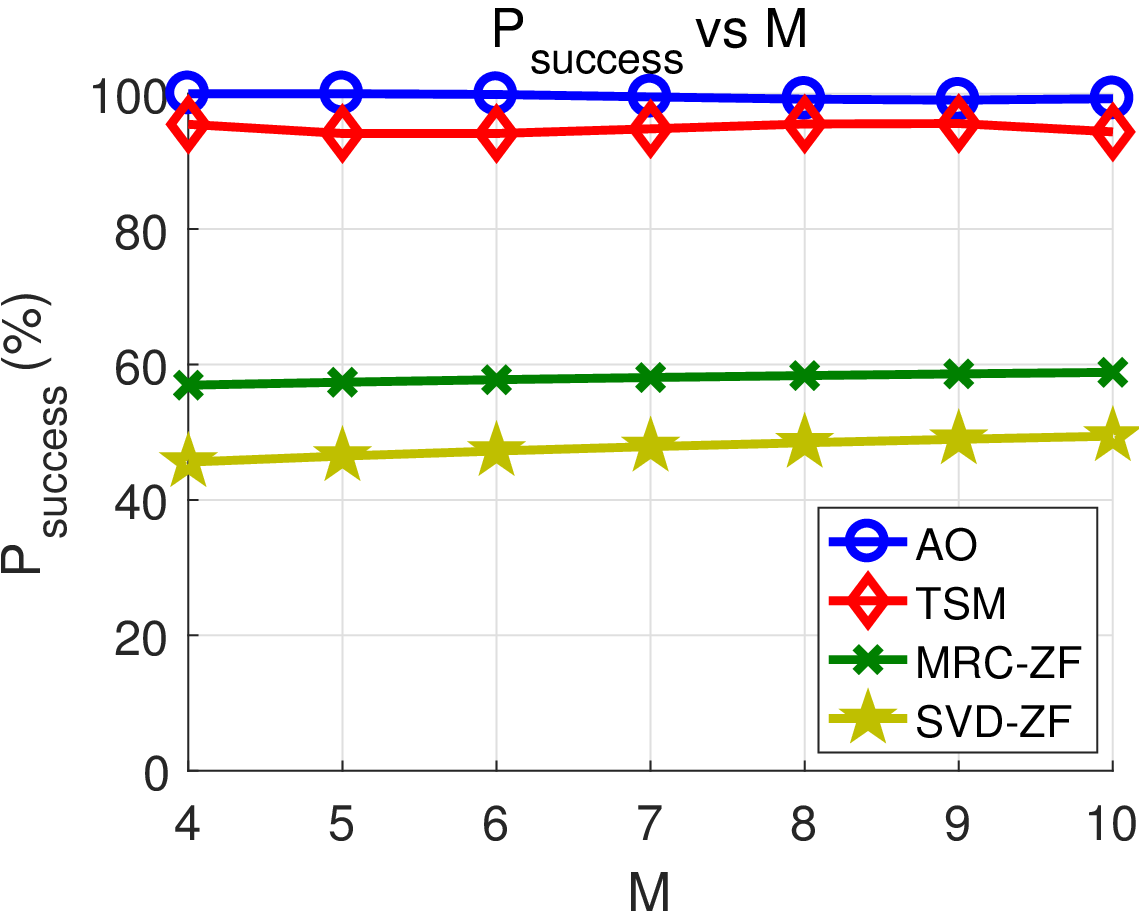}}}
\caption{$P$ and $P_{\text{success}}$ vs $M$. $(K, N, L)=(4, 4, 4), \: \gamma=5$ dB, $\gamma_{\text{ch}}=0.01$.}
\end{figure}          
\vspace{-7mm}
In Fig. 7, we see the effect of the number of CP antennas $M$. The main observation is that the performance enhancement obtained by increasing $M$ is very limited. Adding an extra antenna to CP mainly affects the power spent in fronthaul transmissions. In our channel model, CP-to-RRH channels are better than RRH-to-MS channels in terms of path-loss, antenna gains and receiver characteristics. This is due to the fact that RRHs are stationary and one can place them by optimizing the corresponding fronthaul channel conditions. Therefore, the portion of $P_{\text{CP}}$ in the total power $P$ is small in general and hence the effect of $M$ on the performance is small compared to the effects of $N$ and $L$. 
\vspace{-4mm}
{\renewcommand{\arraystretch}{0.7}
\begin{table}[H]
\caption{$P$ values in dBW for various quadruples of $(K, N, L, M)$ for $\gamma=5$ dB and $\gamma_{\text{ch}}=0.01$}
\centering
\begin{tabular}{| c | c | c | c | c | c | c | c | c | c |}
\hline
$K$ & $N$ & $L$ & $M$ & $P$ (AO) & $P$ (TSM) & $P$ (MRC-ZF) & $P$ (SVD-ZF) & $P$ (Bound) & $P$ (AO) $\: - \: P$ (Bound) \\
\hline
$2$ & $2$ & $4$ & $4$ & $27.52$ & $28.15$ & $33.3$ & $33.61$ & $24.43$ & $3.11$ \\
\hline
$3$ & $2$ & $4$ & $6$ & $27.15$ & $29.42$ & $30.26$ & $31.23$ & $23.82$ & $3.33$ \\
\hline
$4$ & $2$ & $4$ & $8$ & $27.5$ & $29.47$ & $32.41$ & $36.49$ & $24.03$ & $3.47$ \\
\hline
$3$ & $3$ & $4$ & $4$ & $24.07$ & $27.02$ & $30.67$ & $31.32$ & $20.54$ & $3.53$ \\
\hline
$4$ & $4$ & $4$ & $4$ & $25.69$ & $26.23$ & $35.31$ & $36.62$ & $20.48$ & $5.21$ \\
\hline
$3$ & $4$ & $3$ & $4$ & $23.73$ & $25.54$ & $30.44$ & $32.83$ & $19.7$ & $4.03$ \\
\hline
$2$ & $4$ & $2$ & $4$ & $23.59$ & $25.11$ & $31.4$ & $33.33$ & $21.28$ & $2.31$ \\
\hline
\end{tabular}
\end{table}
}
\vspace{-5mm}

In Table II, we compare the performances when the ratios $\dfrac{K}{M}, \dfrac{K}{N}, \dfrac{K}{L}$ are fixed. The first three rows show the cases where $\dfrac{K}{M}, N, L$ are fixed; the first, fourth and fifth rows are related to the case where $\dfrac{K}{N}, M, L$ are fixed, and finally the last three rows correspond to the case where $\dfrac{K}{L}, M, N$ are fixed. We observe that for each three cases, the performance loss of the best method AO compared to the bound is an increasing function of the number of users $K$. This is due to the fact that achieving bound requires perfect elimination of the interference due to undesired users which becomes harder as the number of users increases. 
\vspace{-4mm}
\begin{figure}[H]
\centering
\captionsetup{justification=centering}
\subfloat{{\includegraphics[width=0.30\textwidth]{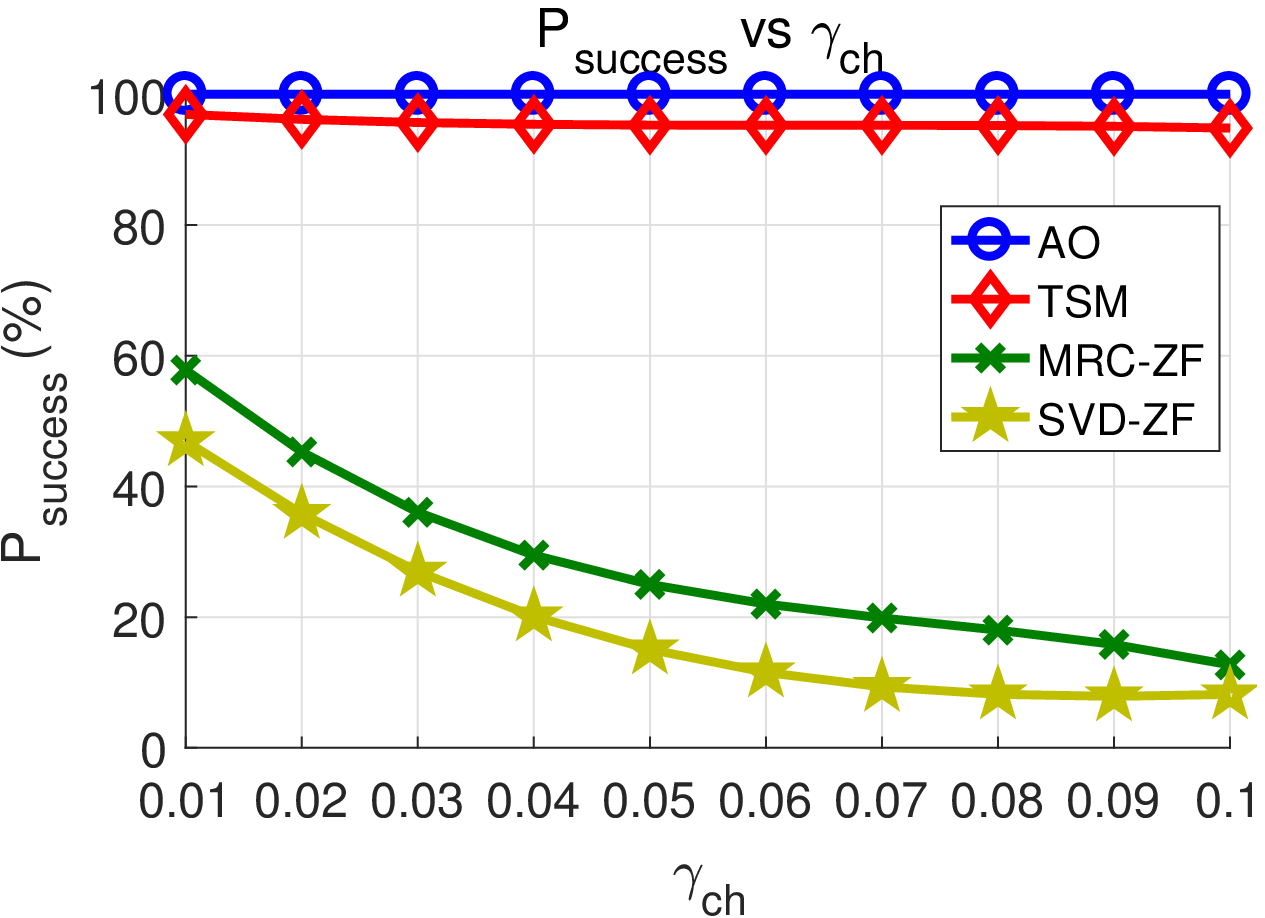}}}
\qquad
\subfloat{{\includegraphics[width=0.27\textwidth]{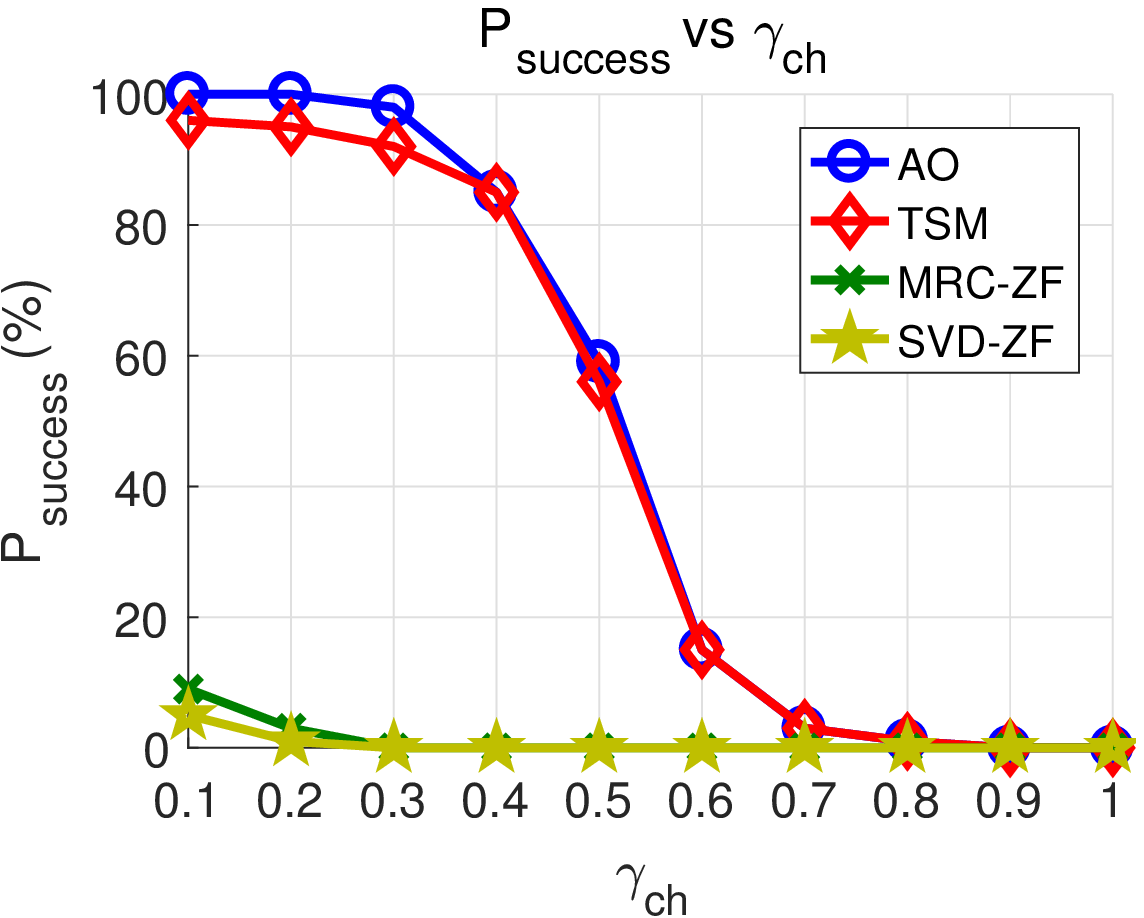}}}
\caption{$P_{\text{success}}$ vs $\gamma_{\text{ch}}$. $(K, N, L, M)=(4, 4, 4, 8), \: \gamma=5$ dB.}
\end{figure}
\vspace{-7mm}
Fig. 8 presents the feasibility ratios with respect to the channel estimation error quality. We observe that if the channel estimation error is large enough, all methods completely fail in the design process. We conclude that convex optimization based methods are more robust to channel errors compared to methods adapted from known beamforming algorithms.
\vspace{-4mm}
\begin{figure}[H]
\centering
\captionsetup{justification=centering}
\includegraphics[width=0.30\textwidth]{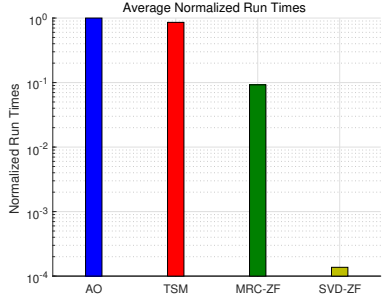}
\caption{Average Complexity Comparison.}
\end{figure}     
\vspace{-7mm}
Fig. 9 shows the normalized average run-times for all methods. Here we take the average over all previously described simulations. We observe that complexity is high for convex optimization based methods. The average run-time of AO is slightly larger than that of TSM. Among all methods we consider, SVD-ZF is the less complex one since it directly finds the solution (if feasible) by solving a linear matrix equation without any solver. On the other hand, its performance is generally not satisfactory in most of the cases. 

In the second part of simulations, we observe the power allocation of users, power sharing between fronthaul and access links, and effect of different user SINR thresholds. We consider two scenarios where RRH and MS locations are fixed. In the both cases, there are a CP with $4$ antennas, $2$ RRHs each with $4$ antennas and $4$ MSs. We only consider AO method to present the results. The first scenario includes various RRH-to-MS distances and second one considers a symmetric placement. In Fig. 10, we present the RRH and MS placements of the two scenarios. 
\vspace{-4mm}
\begin{figure}[ht]
\centering
\captionsetup{justification=centering}
\subfloat{{\includegraphics[width=0.31\textwidth]{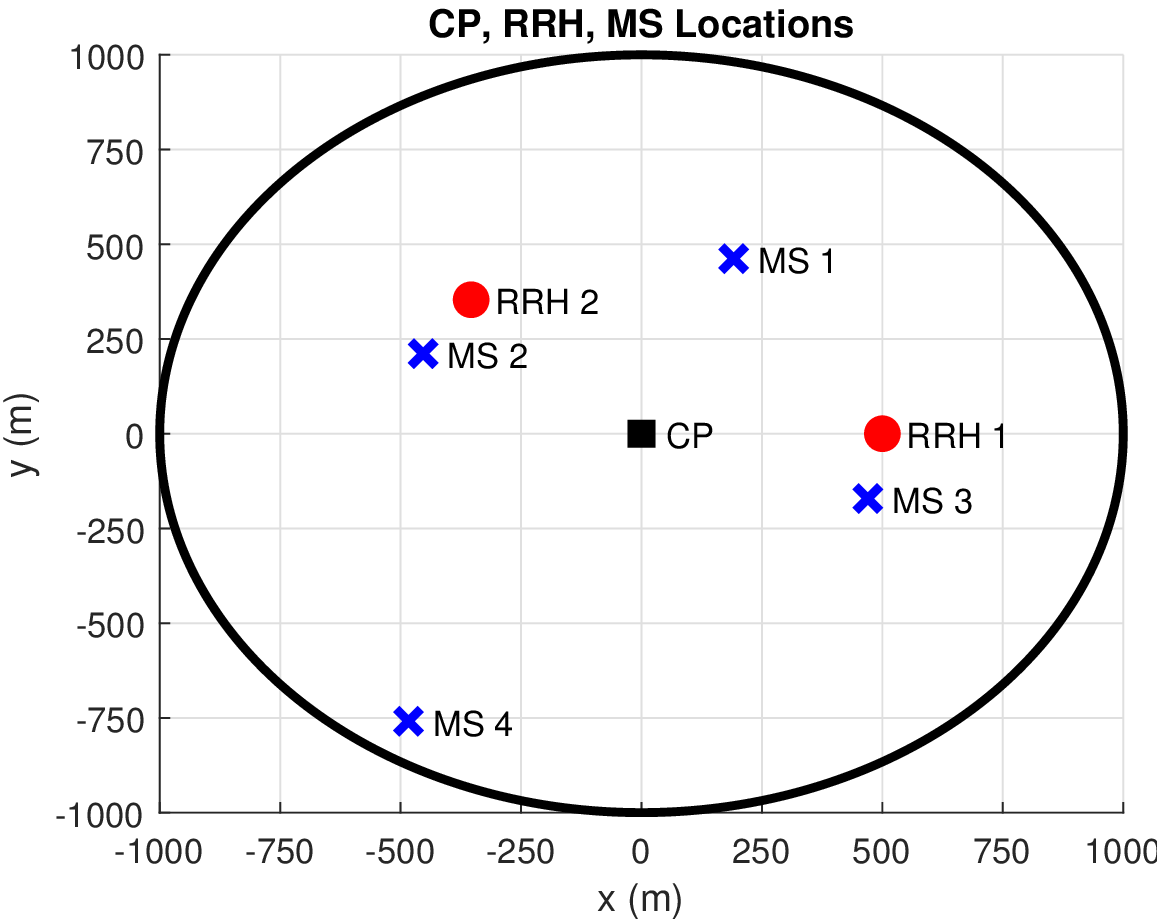}}}
\qquad
\subfloat{{\includegraphics[width=0.31\textwidth]{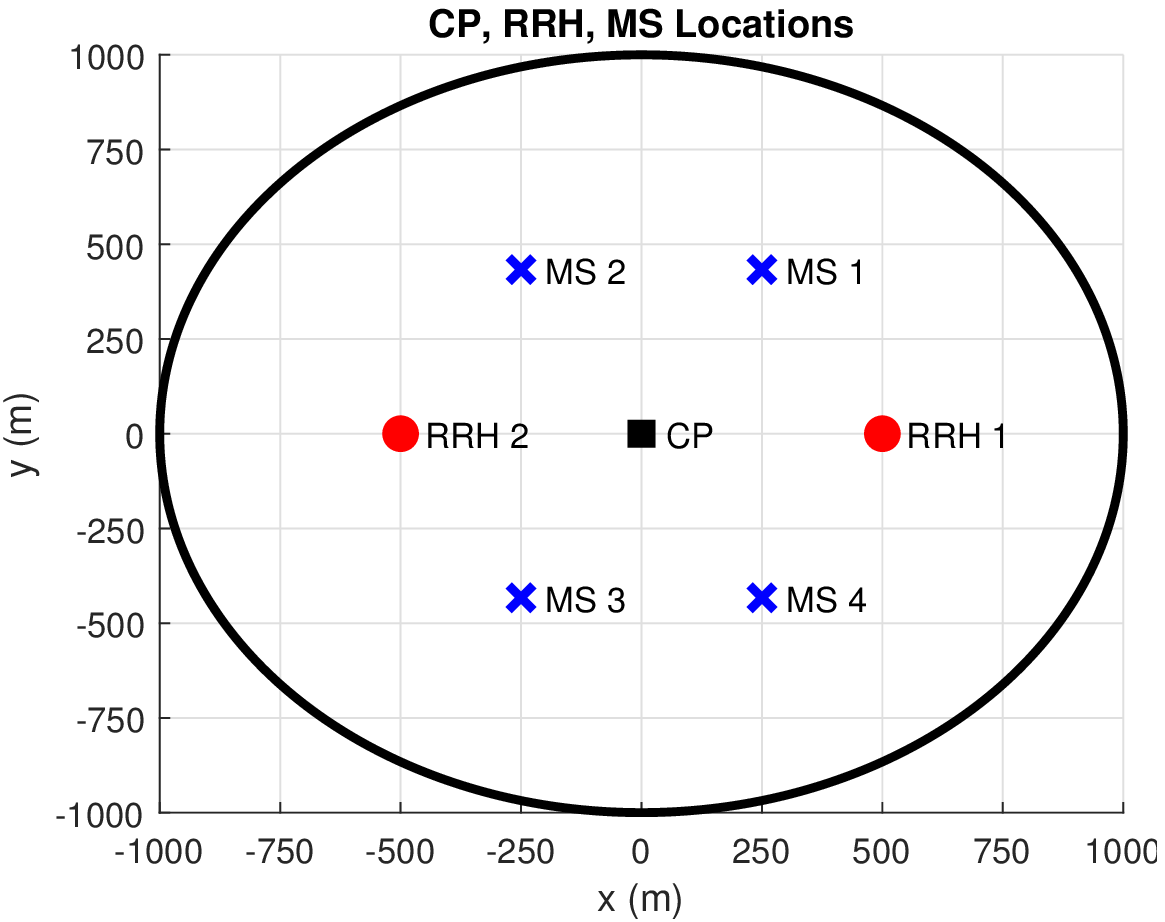}}}
\caption{Scenario 1 (left) and Scenario 2 (right) RRH and MS placements.}
\end{figure}
\vspace{-7mm}
To present the power allocation of users for both fronthaul and access links, we define 
\begin{equation} \label{Pow_alloc1}
\begin{aligned}
P_{\text{CP},k}&=\textbf{v}_k^H\textbf{v}_k, \: P_{\text{RRH},k}=\textbf{v}_k^H \left(\widehat{\textbf{G}}\textbf{W}^H\textbf{W}\widehat{\textbf{G}}^H + \tr\left(\textbf{W}^H \textbf{W} \bm{\Sigma}_1\right)\textbf{I}_M\right) \textbf{v}_k, \\
P_{\text{RRH,amp-noise},k}&=\dfrac{1}{K}\sigma_{\text{RRH}}^2\tr(\textbf{W}^H\textbf{W}), \: \forall k
\end{aligned}
\end{equation}
where $P_{\text{CP},k}, P_{\text{RRH},k}, P_{\text{RRH,amp-noise},k}$ are the fronthaul link power, access link power and RRH amplified noise power for $k$-th user. Notice that we have 
\vspace{-4mm}
\begin{equation} \label{Pow_alloc2}
P_{\text{CP}}=\displaystyle\sum_{k=1}^K P_{\text{CP},k}, \: P_{\text{RRH}} = \displaystyle\sum_{k=1}^K \left(P_{\text{RRH},k} + P_{\text{RRH,amp-noise},k}\right).
\end{equation}
We know that RRH receiver noise is amplified and forwarded to users in AF type relaying. The related term is given in (\ref{r_k_2}) as the first part of the noise term. We equally divide RRH amplified noise power between users as shown in (\ref{Pow_alloc1}). 
\vspace{-4mm}
\begin{figure}[H]
\centering
\captionsetup{justification=centering}
\subfloat{{\includegraphics[width=0.32\textwidth]{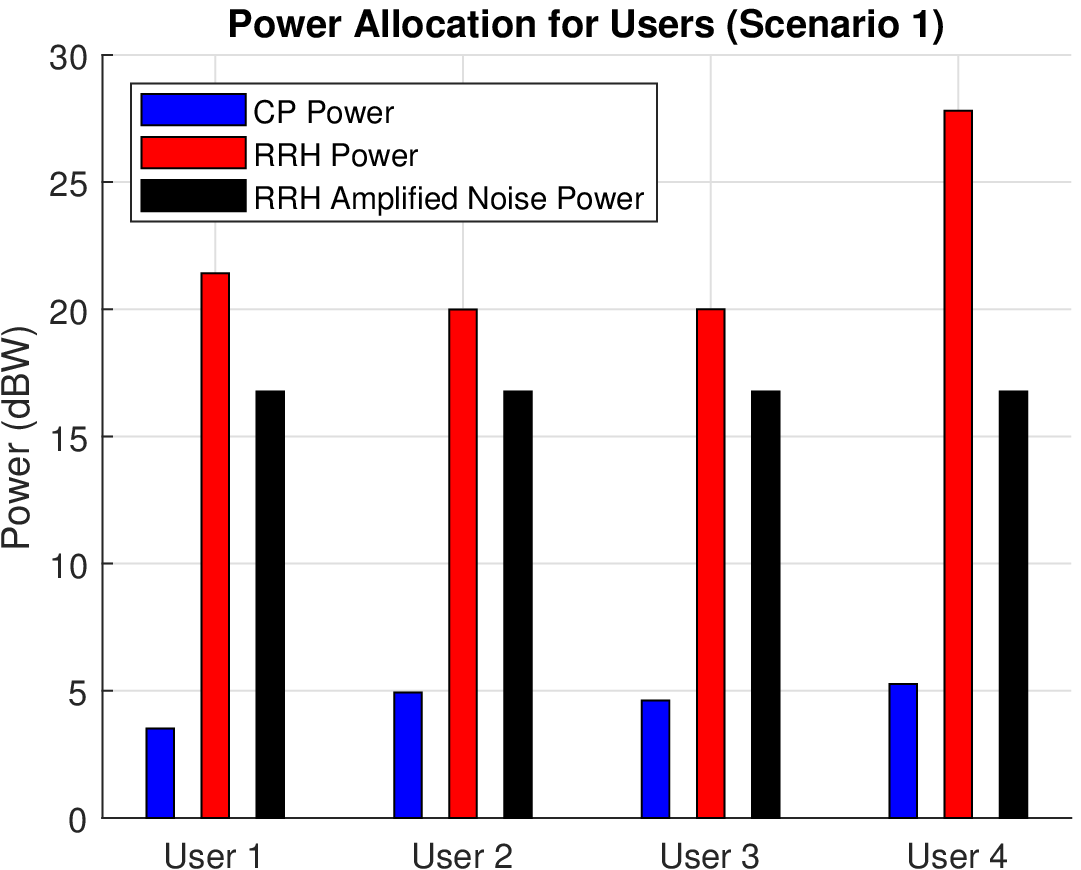}}}
\qquad
\subfloat{{\includegraphics[width=0.32\textwidth]{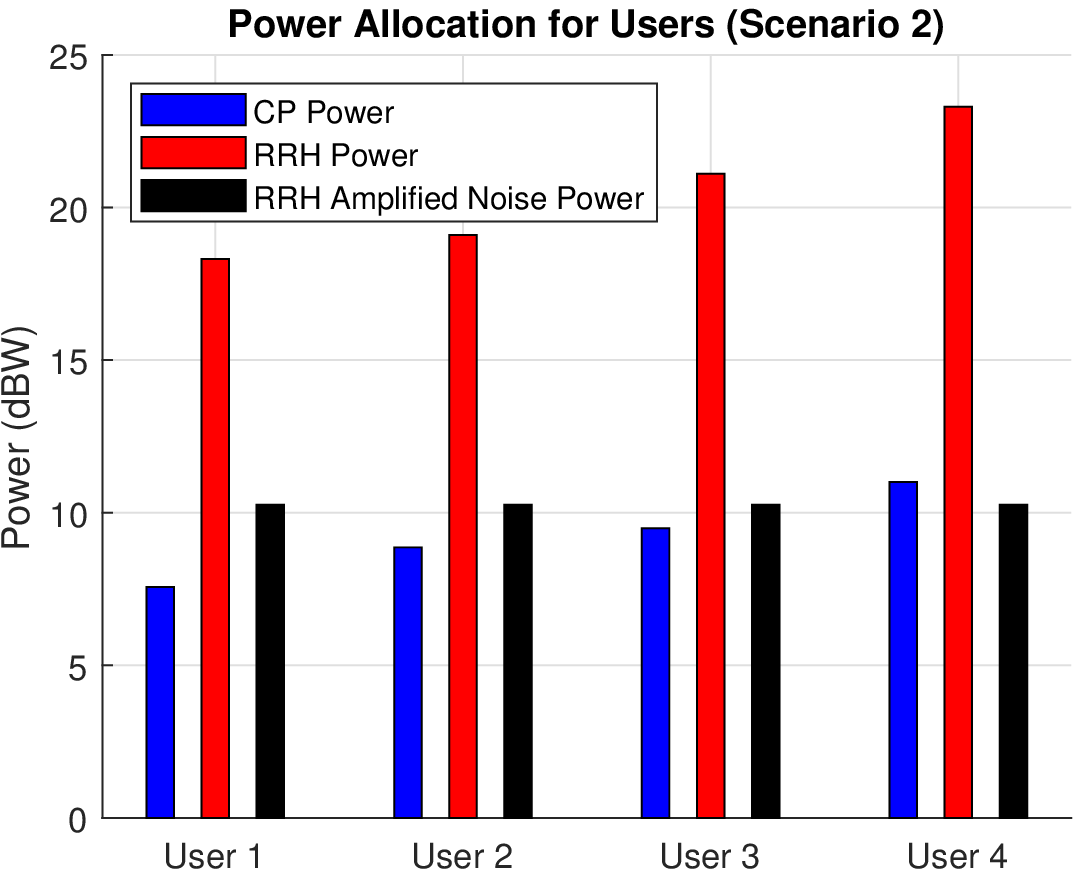}}}
\caption{Power Allocations for Scenario 1 (left) and Scenario 2 (right).}
\end{figure}
\vspace{-7mm}
In the left part of Fig. 11, we observe the power allocation of users for Scenario 1. We take equal SINR thresholds $\gamma_k=\gamma=5$ dB for all users. Notice that fronthaul powers are smaller compared to access link powers. This is due to the path-loss and antenna gain model that we use. As CP and RRHs are stationary, we assume that one can optimize the locations of CP and RRHs so that the corresponding channel conditions are good. We also assume that CP antenna array design is more flexible compared to RRH and MS equipments, and hence we use higher gain antennas for CP. We also observe that $P_{\text{RRH},4} > P_{\text{RRH},1}>P_{\text{RRH},2} \approx P_{\text{RRH},3}$. This is expected considering the locations of users. The distance between MS $4$ and both two RRHs is large and hence it requires the largest power. On the other hand, since MS $2$ and $3$ are close to some RRH, they require the smallest power. MS $1$ distance to both RRHs is at intermediate level and hence the corresponding power is in between the other three MSs. As a final remark, we observe that RRH amplified noise powers are significantly large and this shows that a well-optimized network design is needed to obtain sufficiently large user SINRs for AF type relaying. 

We present the power allocation of users for Scenario 2 in the right part of Fig. 11. In this case, we use a symmetric placement of RRHs and MSs and consider the effect of different user SINR thresholds by taking $\gamma_1=4, \gamma_2=6, \gamma_3=8, \gamma_4=10$ dB. We observe that as the SINR threshold increases, the corresponding user power of both fronthaul and access links also increases. The operator can adjust the user SINR thresholds according to the priority of users. The power required to serve a more prior user will be larger as also presented in this example scenario.  

\vspace{-5mm}
\section{Conclusions}
In this study, we analyzed the joint beamformer design problem in downlink C-RAN with wireless fronthaul. We considered the case where AF type relaying is used in RRHs without the capability of baseband processing. We assumed that channel coefficients are available with some additive error with known second order statistics. We derived a novel theoretical lower bound for the total power spent under SINR constraints. We proposed two convex optimization based methods and two other methods adapted from known beamforming strategies to observe the tightness of the bound. We have shown that first two methods have better performances but their complexities are also higher. In general, the performance of the best method is close to the bound and the difference is less than $1$ dB for some cases. The results show the effectiveness of the bound as well as the performances of various solution techniques. For C-RAN systems, there are other beamforming design techniques that are not analyzed in this study but studied in the literature. We have found at least one method performing close to the bound and this is enough to show the tightness of the bound proposed. 

As a future work, the approach used in this study to derive a performance bound can be adapted to DF and DCF based relaying and also to full-duplex RRH case. In all simulations, we observed that SDR based methods always produce rank-1 results. This fact can be proved in a future study. Finally one can search the necessary conditions required for the equality case of the bound to gain insight about the optimal algorithm.  
\vspace{-3mm}
\appendices
\section{Achievability of Rate}
We use the idea given in \cite{InformationTheory} to show that the rate $\log_2(1+\text{SINR}_k)$ is achievable for $k$-th user where $\text{SINR}_k$ is defined by (\ref{SINR_2}). We find a lower bound to the mutual information $I(r_k; s_k)$ between the received signal $r_k$ and the information signal $s_k$. Using the facts that conditioning decreases entropy $h(\cdot)$, the entropy is maximized for Gaussian distribution when the variance is fixed, the entropy is invariant under translation, and $s_k$ and $r_k$ are zero-mean, we can write
\begin{equation} \label{App1}
\begin{aligned}
I(r_k; s_k)&=h(s_k)-h(s_k | r_k) = h(s_k)-h(s_k-\alpha r_k | r_k) \geq h(s_k)-h(s_k-\alpha r_k) \\
&\geq \log\left(\pi e \mathbb{E}\left[|s_k|^2\right]\right) - \log\left(\pi e \mathbb{E}\left[|s_k-\alpha r_k|^2\right]\right) = \log\left(\dfrac{\mathbb{E}\left[|s_k|^2\right]}{\mathbb{E}\left[|s_k-\alpha r_k|^2\right]}\right).
\end{aligned}
\end{equation}
Here we assume that $s_k$ is complex Gaussian and $\alpha$ is any complex constant. (\ref{App1}) is true for any $\alpha$ and specifically we choose $\alpha=\mathbb{E}\left[r_k^{*}s_k\right]/\mathbb{E}\left[|r_k|^2\right]$ to get
\vspace{-5mm}
\begin{equation} \label{App2}
I(r_k; s_k) \geq \log\left(1 + \dfrac{|\mathbb{E}\left[r_k^{*}s_k\right]|^2}{\mathbb{E}\left[|r_k|^2\right] \cdot \mathbb{E}\left[|s_k|^2\right] - |\mathbb{E}\left[r_k^{*}s_k\right]|^2}\right).
\end{equation}

\noindent Using the equation of $r_k$ in (\ref{r_k_2}) and the fact $\mathbb{E}\left[|s_k|^2\right]=1$, we obtain that $|\mathbb{E}\left[r_k^{*}s_k\right]|^2 = P_d$ and $\mathbb{E}\left[|r_k|^2\right] = P_d + P_{I,1}+P_{I,2}+P_n$ where $P_d, P_{I,1}, P_{I,2}, P_n$ are defined in (\ref{P_d}). Therefore we conclude that $I(r_k; s_k)$ is at least $\log_2\left(1+\dfrac{P_d}{P_{I,1}+P_{I,2}+P_n}\right) = \log_2(1+\text{SINR}_k)$ bits.

\section{Proof of (\ref{B_bound})}
Using (\ref{B5}) and (\ref{B7}), we get
\begin{equation} \label{B9}
x_1 \geq c_1\left[c_2x_2 + (c_3x_5+c_4)\left(x_3+\dfrac{c_2}{d_1}x_3\right)+c_5\right], \: \:
y \geq x_2+x_5+(c_3x_5+c_6)\dfrac{x_3}{d_1}.
\end{equation}
(\ref{B6}) and (\ref{B9}) yields
\begin{equation} \label{B10}
(d_1-c_1c_2)x_2 \geq c_1\left[(c_3x_5+c_4)\left(x_3+\dfrac{c_2}{d_1}x_3\right)+c_5\right]
\end{equation}
and (\ref{B8}) and (\ref{B9}) yields
\begin{equation} \label{B11}
\left[\dfrac{x_5d_2}{c_1}-(c_3x_5+c_4)\left(1+\dfrac{c_2}{d_1}\right)\right]x_3 \geq c_2x_2+c_5.
\end{equation}
(\ref{B10}) implies that $d_1>c_1c_2$. By (\ref{B10}), (\ref{B11}) and some simplifications, we obtain that
\begin{equation} \label{B12}
x_3 \geq \dfrac{d_1c_5}{\left(\dfrac{d_1d_2}{c_1}-c_2d_2-c_3d_1-c_2c_3\right)x_5-c_4(d_1+c_2)}
\end{equation}
and the denominator in (\ref{B12}) should be positive. Using (\ref{B10}) and (\ref{B12}) we get
\begin{equation} \label{B13}
x_2 \geq \dfrac{d_2c_5}{\left(\dfrac{d_1d_2}{c_1}-c_2d_2-c_3d_1-c_2c_3\right)x_5-c_4(d_1+c_2)}.
\end{equation}
Using (\ref{B9}), (\ref{B12}) and (\ref{B13}) we find that
\begin{equation} \label{B14}
y \geq x_5+\dfrac{d_2c_5+c_5(c_3x_5+c_6)}{\left(\dfrac{d_1d_2}{c_1}-c_2d_2-c_3d_1-c_2c_3\right)x_5-c_4(d_1+c_2)}.
\end{equation}
Define $x=ax_5-b$ where $a=\dfrac{d_1d_2}{c_1}-c_2d_2-c_3d_1-c_2c_3, \: b=c_4(d_1+c_2)$. Since $x$ is the denominator of (\ref{B12}), it is positive. As $x_5$ and $b$ are positive, we conclude that $a$ is also positive. We can write (\ref{B14}) in terms of $x$ as 
\begin{equation} \label{B15}
y \geq \dfrac{1}{a}\left(b+c_3c_5+x+\dfrac{c_3c_5b+c_5(d_2+c_6)a}{x}\right).
\end{equation}
Finally, using (\ref{B15}) and Arithmetic-Geometric Mean Inequality, we get the desired result in (\ref{B_bound}).


\begin{thebibliography}{100}
\bibitem{C-RAN} O. Simeone, A. Maeder, M. Peng, O. Sahin and W. Yu, ``Cloud radio access network: Virtualizing wireless access for dense heterogeneous systems," in \textit{Journal of Comm. and Networks}, vol. 18, no. 2, pp. 135-149, April 2016.

\bibitem{AF-C-RAN} C. Kuo, S. Wu and C. Tseng, ``Robust Linear Beamformer Designs for Coordinated Multi-Point AF Relaying in Downlink Multi-Cell Networks," in \textit{IEEE Trans. on Wireless Comm.}, vol. 11, no. 9, pp. 3272-3283, September 2012.

\bibitem{AF-Relay1} C. Wang, X. Dong and Y. Shi, ``Robust relay design for two-way multi-antenna relay systems with imperfect CSI," in \textit{Journal of Comm. and Networks}, vol. 16, no. 1, pp. 45-55, Feb. 2014.

\bibitem{AF-Relay2} B. K. Chalise and L. Vandendorpe, ``Optimization of MIMO Relays for Multipoint-to-Multipoint Comm.: Nonrobust and Robust Designs," in \textit{IEEE Trans. on Signal Process.}, vol. 58, no. 12, pp. 6355-6368, Dec. 2010.

\bibitem{DF-1} B. Hu, C. Hua, J. Zhang, C. Chen and X. Guan, ``Joint Fronthaul Multicast Beamforming and User-Centric Clustering in Downlink C-RANs," in \textit{IEEE Trans. on Wireless Comm.}, vol. 16, no. 8, pp. 5395-5409, Aug. 2017.

\bibitem{DF-2} B. Hu, C. Hua, C. Chen and X. Guan, ``Joint Beamformer Design for Wireless Fronthaul and Access Links in C-RANs," in \textit{IEEE Trans. on Wireless Comm.}, vol. 17, no. 5, pp. 2869-2881, May 2018.

\bibitem{DCF-1} S. Park, K. Lee, C. Song and I. Lee, ``Joint Design of Fronthaul and Access Links for C-RAN With Wireless Fronthauling," in \textit{IEEE Signal Process. Lett.}, vol. 23, no. 11, pp. 1657-1661, Nov. 2016.

\bibitem{DCF-2} S. Park, C. Song and K. Lee, ``Inter-Cluster Design of Wireless Fronthaul and Access Links for the Downlink of C-RAN," in \textit{IEEE Wireless Comm. Lett.}, vol. 6, no. 2, pp. 270-273, April 2017.

\bibitem{Ch-Add-1} H. Du and P. Chung, ``A Probabilistic Approach for Robust Leakage-Based MU-MIMO Downlink Beamforming with Imperfect Channel State Information," in \textit{IEEE Trans. on Wireless Comm.}, vol. 11, no. 3, pp. 1239-1247, March 2012.

\bibitem{Ch-Add-2} D. Wang, Y. Wang, R. Sun and X. Zhang, ``Robust C-RAN Precoder Design for Wireless Fronthaul with Imperfect Channel State Information," \textit{2017 IEEE Wireless Comm. and Netw. Conf. (WCNC)}, San Francisco, CA, 2017, pp. 1-6.

\bibitem{Ch-Add-3} D. Yan, R. Wang, E. Liu and Q. Hou, ``ADMM-Based Robust Beamforming Design for Downlink Cloud Radio Access Networks," in \textit{IEEE Access}, vol. 6, pp. 27912-27922, 2018.

\bibitem{AF-Relay3} J. Li and M. Haardt, ``Robust MIMO Relay Precoder Design for Multiple Operator One-Way Relaying with Imperfect Channel State Information," \textit{International Symposium on Wireless Commun. Systems}, Ilmenau, Germany, 2013, pp. 1-5.

\bibitem{AF-Relay4} B. K. Chalise and L. Vandendorpe, ``MIMO Relay Design for Multipoint-to-Multipoint Comm. With Imperfect Channel State Information," in \textit{IEEE Trans. on Signal Process.}, vol. 57, no. 7, pp. 2785-2796, July 2009.

\bibitem{AF-Relay5-ch-err} P. Ubaidulla and A. Chockalingam, ``Relay Precoder Optimization in MIMO-Relay Networks With Imperfect CSI," in \textit{IEEE Trans. on Signal Process.}, vol. 59, no. 11, pp. 5473-5484, Nov. 2011.

\bibitem{Wired-Rate1} J. Zhang, R. Chen, J. G. Andrews, A. Ghosh and R. W. Heath, ``Networked MIMO with clustered linear precoding," in \textit{IEEE Trans. on Wireless Comm.}, vol. 8, no. 4, pp. 1910-1921, April 2009.

\bibitem{Wired-Rate2} A. Liu and V. K. N. Lau, ``Joint Power and Antenna Selection Optimization in Large Cloud Radio Access Networks," in \textit{IEEE Trans. on Signal Process.}, vol. 62, no. 5, pp. 1319-1328, March 2014.

\bibitem{Wired-Rate3} B. Dai and W. Yu, ``Sparse Beamforming and User-Centric Clustering for Downlink Cloud Radio Access Network," in \textit{IEEE Access}, vol. 2, pp. 1326-1339, 2014.

\bibitem{Wired-UDD} Y. Huang et al., ``Distributed Multicell Beamforming Design Approaching Pareto Boundary with Max-Min Fairness," in \textit{IEEE Trans. on Wireless Comm.}, vol. 11, no. 8, pp. 2921-2933, August 2012.

\bibitem{Wired-LimitedFronthaul} R. Zakhour and D. Gesbert, ``Optimized Data Sharing in Multicell MIMO With Finite Backhaul Capacity," in \textit{IEEE Trans. on Signal Process.}, vol. 59, no. 12, pp. 6102-6111, Dec. 2011.

\bibitem{Wired-SDR} F. Zhuang and V. K. N. Lau, ``Backhaul Limited Asymmetric Cooperation for MIMO Cellular Networks via Semidefinite Relaxation," in \textit{IEEE Trans. on Signal Process.}, vol. 62, no. 3, pp. 684-693, Feb.1, 2014.

\bibitem{Wired-UserMax} H. Wai and W. Ma, ``A decentralized method for joint admission control and beamforming in coordinated multicell downlink," \textit{Asilomar Conf. on Signals, Systems and Computers}, Pacific Grove, CA, 2012, pp. 559-563.

\bibitem{Wired-GreenCRAN} Y. Shi, J. Zhang and K. B. Letaief, ``Group Sparse Beamforming for Green Cloud-RAN," in \textit{IEEE Trans. on Wireless Comm.}, vol. 13, no. 5, pp. 2809-2823, May 2014.

\bibitem{Wired-ZF} A. Chowdhery, W. Yu and J. M. Cioffi, ``Cooperative Wireless Multicell OFDMA Network with Backhaul Capacity Constraints," \textit{2011 IEEE International Conf. on Comm. (ICC)}, Kyoto, 2011, pp. 1-6.

\bibitem{Wired-Heuristic1} J. Zhao, T. Q. S. Quek and Z. Lei, ``Coordinated Multipoint Transmission with Limited Backhaul Data Transfer," in \textit{IEEE Trans. on Wireless Comm.}, vol. 12, no. 6, pp. 2762-2775, June 2013.

\bibitem{Wired-Heuristic2} A. Papadogiannis, D. Gesbert and E. Hardouin, ``A Dynamic Clustering Approach in Wireless Networks with Multi-Cell Cooperative Process.," \textit{2008 IEEE International Conf. on Comm.}, Beijing, 2008, pp. 4033-4037.

\bibitem{Wired-Heuristic3} F. E. Kadan and A. Ö. Yılmaz, ``Optimized asymmetric cooperation for downlink cloud radio access network under per-base station data transfer constraint," \textit{2017 IEEE International Conf. on Comm. Workshops}, Paris, 2017, pp. 132-137.

\bibitem{Wired-Imperfect-CSI} Y. Shi, J. Zhang and K. B. Letaief, ``Robust Group Sparse Beamforming for Multicast Green Cloud-RAN With Imperfect CSI," in \textit{IEEE Trans. on Signal Process.}, vol. 63, no. 17, pp. 4647-4659, Sept., 2015.

\bibitem{Wired-ClusterFormation} Z. Zhao, M. Peng, Z. Ding, C. Wang and H. V. Poor, ``Cluster formation in cloud-radio access networks: Performance analysis and algorithms design," \textit{2015 IEEE International Conf. on Comm. (ICC)}, London, 2015, pp. 3903-3908.

\bibitem{Wired-Delay} Jian Li, Mugen Peng, Aolin Cheng and Yuling Yu, ``Delay-aware cooperative multipoint transmission with backhaul limitation in cloud-RAN," \textit{2014 IEEE International Conf. on Comm. Workshops (ICC)}, Sydney, NSW, 2014, pp. 665-670.

\bibitem{AF-Relay6} A. Papadogiannis, A. G. Burr and M. Tao, ``On the Maximum Achievable Sum-Rate of Interfering Two-Way Relay Channels," in \textit{IEEE Comm. Lett.}, vol. 16, no. 1, pp. 72-75, January 2012.

\bibitem{AF-Relay7} D. Ponukumati, F. Gao and L. Fan, ``Robust General Rank Precoding Design for Amplify-and-Forward Relay Network," \textit{2010 IEEE Global TeleComm. Conf. GLOBECOM 2010}, Miami, FL, 2010, pp. 1-5.

\bibitem{Von-Neumann} A. W. Marshall, I. Olkin and B. Arnold, \textit{Inequalities: Theory of Majorization and Its Applications}, 2nd ed. New York: Springer, 2011, pp. 340-341.

\bibitem{CS} J. M. Steele, \textit{The Cauchy-Schwarz Master Class: An Introduction to the Art of Mathematical Inequalities}, Cambridge, Cambridge University Press, 2004.

\bibitem{AM-GM} K. M. Chong, ``The Arithmetic Mean-Geometric Mean Inequality: A New Proof," \textit{Mathematics Magazine}, vol. 49, no. 2, pp. 87-88, March 1976.

\bibitem{vec-eqn} R. A. Horn and C. R. Johnson, \textit{Topics in Matrix Analysis}. Cambridge, U.K.: Cambridge Univ. Press, 1991.

\bibitem{SeDuMi} J. F. Sturm, ``Using SeDuMi 1.02, a MATLAB toolbox for optimization
over symmetric cones," \textit{Optimization Methods Software}, vol. 11-12, pp. 625–653, 1999.

\bibitem{CVX} M. Grant and S. Boyd. (Jan. 2020). \textit{CVX: MATLAB Software for
Disciplined Convex Programming}, Version 2.2. [Online]. Available: http://cvxr.com/cvx

\bibitem{Mosek} MOSEK ApS. (2020). \textit{The MOSEK Optimization Software}. Version 9.2. [Online]. Available: http://www.mosek.com

\bibitem{MCT} R. G. Bartle and D. R. Sherbert, \textit{Introduction to Real Analysis}, 4th Edition. Hoboken, NJ: Wiley, 2011, pp. 71–72.

\bibitem{3GPP} \textit{Further Advancements for E-UTRA Physical Layer Aspects (Release 9)}, document 3GPP TR 36.814, March 2010.

\bibitem{InformationTheory} M. Medard, ``The effect upon channel capacity in wireless communications of perfect and imperfect knowledge of the channel," in \textit{IEEE Trans. on Inf. Theory}, vol. 46, no. 3, pp. 933-946, May 2000.

\end{thebibliography}
\end{document}